**Manuscript Type:** Original Research Article

**Title:** Bias in studies of prenatal exposures using real-world data due to pregnancy identification method

**Authors:** Chase D. Latour,[1] Jessie K. Edwards,[1,2] Michele Jonsson Funk,[1] Elizabeth A. Suarez,[3,4] Kim Boggess,[5] Mollie E. Wood[1]

**Affiliations:**
[1] Department of Epidemiology, University of North Carolina at Chapel Hill, Chapel Hill, NC, USA
[2] Carolina Population Center, University of North Carolina at Chapel Hill, Chapel Hill, NC, USA
[3] Center for Pharmacoepidemiology and Treatment Science, Rutgers Institute for Health, Health Care Policy and Aging Research, New Brunswick, NJ, USA
[4] Department of Biostatistics and Epidemiology, Rutgers School of Public Health, Piscataway, NJ, USA
[5] Department of Obstetrics and Gynecology, University of North Carolina Chapel Hill, Chapel Hill, NC, USA

**Running Head:** Bias Due to Pregnancy Identification in Real-World Data

**Conflicts of Interest:** CDL has received payment from Amgen, Regeneron Pharmaceuticals, and Target RWE for unrelated work. Abbvie, Astellas, Boehringer Ingelheim, GlaxoSmithKline (GSK), Takeda, Sarepta, and UCB Bioscience have collaborative agreements with the Center for Pharmacoepidemiology housed in the Department of Epidemiology which provides salary support to MJF. MJF is a member of the Scientific Steering Committee (SSC) of a post-approval safety study of an unrelated drug class funded by GSK. All compensation for services provided on the SSC is invoiced by and paid to UNC Chapel Hill. MEW is affiliated with the Center for Pharmacoepidemiology at the University of North Carolina at Chapel Hill and provides limited methods consulting to Center members unrelated to the present work. JKE, EAS, and KB have no other conflicts to disclose.

**Funding:** This research was partially supported by a National Research Service Award Pre-Doctoral/Post-Doctoral Traineeship from the Agency for Healthcare Research and Quality sponsored by The Cecil G. Sheps Center for Health Services Research, The University of North Carolina at Chapel Hill, Grant No. T32-HS000032 (CDL). This research was supported in part by a training grant from the National Institute of Child Health and Development [T32 HD052468] (CDL).

**Data and Computing Code Availability:** All code used to generate and analyze the data presented in this article is available on GitHub: https://github.com/chasedlatour/preg_cohort_construction.

**Abstract**

**Background:** Researchers typically identify pregnancies in healthcare data based on observed outcomes (e.g., delivery). This outcome-based approach misses pregnancies that received prenatal care but whose outcomes were not recorded (e.g., at-home miscarriage), potentially inducing selection bias in effect estimates for prenatal exposures. Alternatively, prenatal encounters can be used to identify pregnancies, including those with unobserved outcomes. However, this prenatal approach requires methods to address missing data.

**Methods:** We simulated 10,000,000 pregnancies and estimated the total effect of initiating treatment on the risk of preeclampsia. We generated data for 36 scenarios in which we varied the effect of treatment on miscarriage and/or preeclampsia; the percentage with missing outcomes (5% or 20%); and the cause of missingness: (1) measured covariates, (2) unobserved miscarriage, and (3) a mix of both. We then created three analytic samples to address missing pregnancy outcomes: observed deliveries, observed deliveries and miscarriages, and all pregnancies. Treatment effects were estimated using non-parametric direct standardization.

**Results:** Risk differences (RDs) and risk ratios (RRs) from the three analytic samples were similarly biased when all missingness was due to unobserved miscarriage (log-transformed RR bias range: -0.12-0.33 among observed deliveries; -0.11-0.32 among observed deliveries and miscarriages; and -0.11-0.32 among all pregnancies). When predictors of missingness were measured, only the all pregnancies approach was unbiased (-0.27-0.33; -0.29-0.03; and -0.02-0.01, respectively).

**Conclusions:** When all missingness was due to miscarriage, the analytic samples returned similar effect estimates. Only among all pregnancies did bias decrease as the proportion of missingness due to measured variables increased.

## 1. Introduction

Estimating the effects of prenatal medication use using administrative healthcare data (e.g., insurance claims, electronic health records) requires identifying pregnancies based on diagnosis or procedure codes that result from contact with the healthcare system.[1] Most pregnancy identification algorithms define pregnancy episodes by first finding codes that indicate that an outcome (e.g., live birth, miscarriage) occurred and then looking for information on gestational age—henceforth the "outcome-based approach".[1,2] The concern with this approach for studies of prenatal exposures is that it conditions on post-exposure events, potentially causing selection bias.[3,4]

Recent efforts to identify pregnancies focus on identifying pregnancies at their first encounter or claim with evidence of pregnancy;[2,5] this approach–henceforth the "prenatal approach—thus minimizes conditioning on post-exposure events. However, when using the prenatal approach, pregnancy outcomes may not be observed for all included pregnancies and missingness may be informative.[2] Administrative healthcare data arise from contact with the healthcare system, and early pregnancy losses are less likely to result in medical encounters than later losses or deliveries.[6]

There has been substantial work focused on bias induced by conditioning on live birth when analyzing data,[7–9] but, to our knowledge, there has been little discussion about bias induced by how pregnancies are identified. In this article, we first illustrate the bias induced by each approach to pregnancy identification using directed acyclic graphs (DAGs). We then use simulation to explore the magnitude and direction of the bias we might expect under different approaches to pregnancy identification and varying extent and sources of missing outcomes. These findings can elucidate scenarios where investigators should be most concerned about bias due to missing pregnancy outcomes.

**2. Description of the Bias**

Figure 1 contains three DAGs for a study of the effect of a prenatal exposure *E* on a multinomial study outcome *D(j)*, where *j* indexes the value of the study outcome: miscarriage, prenatal preeclampsia, and non-preeclamptic delivery (Methods S1).[10–13] Additionally, we include *C*, a risk factor for *D(j)*, and *S*, an indicator for observing the pregnancy outcome. We use a box around *S*=1 to indicate that only individuals with an observed pregnancy outcome are included in the analytic cohort. We follow established rules from the single outcome setting for analyzing DAGs to evaluate bias due to selection and missing data.[14–17]

*"Outcome-Based Approach"*

We frame the bias induced by the "outcome-based approach" as selection bias because observing the pregnancy outcome in the source data determines selection into the study. We enumerate three populations for understanding potential selection bias:[15] target population, study sample, and analytic sample (Table 1). We focus on selection from the study sample to the analytic sample, as differences in their effect estimates represent selection bias.

In Figure 1A, individuals with observed pregnancy outcomes (i.e., *S*=1) are a random sample of all pregnancies. This is an ideal, if unrealistic, scenario in which we expect no bias.

In Figure 1B, observation of the pregnancy outcome is determined by *C*. In this scenario, a complete case analysis can cause bias: individuals within stratum of *C* may be more/less likely to be censored but have a different risk of the outcome. This bias can be overcome by weighting risks according to individuals' probabilities of having their outcome observed dependent upon *C* (i.e., selection probabilities, $P(S = 1|C = c)$, where $c$ indicates their observed value of $C$). However, these probabilities cannot be derived using the "outcome-based approach," as pregnancies without observed outcomes are not identified.

In Figure 1C, observing the outcome is determined by the outcome itself (i.e., $P(S = 1|D(j) = d(j))$, where $d(j)$ represents their observed value of $D(j)$). Because the reason for missingness is unmeasured, this problem cannot be resolved analytically. Investigators must leverage sensitivity analyses to quantify the potential impact of selection bias.[15,16,18] For the "outcome-based approach," assumptions about the degree and direction of selection are based upon data from other populations and clinical expertise.[19]

The sources of selection in Figures 1B and 1C are likely. In the context of electronic health records, severity of a comorbidity may influence referral to another healthcare system and the outcome risk (Figure 1B).[20] Further, someone's true outcome may influence observation (Figure 1C): an early miscarriage that can be managed at home is less likely to be observed than a delivery that requires medical oversight.[6,20] The "outcome-based approach" provides minimal options to address these likely potential sources of bias.

*"Prenatal Approach"*

We frame the bias induced by the "prenatal approach" as an issue of missing data: all pregnancies are identified but some are missing outcomes. We consider again the DAGs from Figure 1, ignoring the box around *S=1* because the analytic sample now includes pregnancies with missing outcomes. Drawing upon prior work,[16,21,22] we can see that the outcomes are

1. missing completely at random (MCAR) in Figure 1A (i.e., individuals with observed outcomes are a true random sample of all pregnancies),
2. missing at random (MAR) in Figure 1B (i.e., observation of a pregnancy outcome is predicted by a measured variable), and
3. missing not at random (MNAR) in Figure 1C (i.e., observation of a pregnancy outcome is predicted by the outcome itself).

A complete case analysis of an analytic sample identified by the "prenatal approach" would introduce selection on *S* and thus produces the same biases as the "outcome-based approach",[16] but more options are available for addressing this bias with a cohort derived via the "prenatal approach". If the data were MAR, we could derive selection probabilities conditional on *C* to overcome the bias. If MNAR, we know the extent of missingness and can quantify the full set of bounds on the treatment effect,[23,24] along with sensitivity analyses that make different assumptions about the missingness mechanisms.[25]

**3. Simulation**

*Description of the Simulation*

We simulated data for a hypothetical observational study based on a recent trial.[26] Our hypothetical study aimed to answer the question: what is the effect of initiating versus not initiating antihypertensives at nine weeks of gestation on the risk of prenatal preeclampsia among pregnant people with chronic hypertension?

*Generating Target Populations*

We generated target populations corresponding to 6 scenarios distinguished by the effect of antihypertensive initiation on the risk of miscarriage (defined as fetal death <20 weeks of gestation; decreased, increased, or no effect) and prenatal preeclampsia (decreased risk or no effect) (Table S1), guided by the DAG in Figure S1.[27]

Step 1. We simulated 10,000,000 pregnancies at 9 weeks of gestation from the last menstrual period and randomly assigned each to have mild (33.3%), moderate (33.3%), or severe (33.3%) chronic hypertension and to reside in a rural (30%) or non-rural (70%) area (Figure S2). We induced confounding by hypertension severity using a Bernoulli random variable to assign treatment according to severity where the probability of initiation was 25% for mild, 50% for moderate, and 75% for severe hypertension. Rurality was also assigned via a Bernoulli random variable but did not affect treatment.

Step 2. We generated weekly potential pregnancy outcomes (i.e., treated and untreated outcomes) at gestational weeks 9-42 (Methods S2). Untreated potential outcomes under low severity disease among non-rural participants mirrored real-world populations.[26,28] Higher hypertension severity and rurality increased risk of miscarriage. The effect of treatment on the weekly risk of miscarriage varied across scenarios from 0.1 to 5 times the untreated weekly risk, dependent upon hypertension severity.

Step 3. We generated weekly potential preeclampsia outcomes at weeks 19-42 (Methods S3). Untreated potential outcomes under mild hypertension mirrored real-world populations.[26,28] Increasing hypertension severity increased risk of preeclampsia, as did residing in a rural area. Treatment affected the weekly risk of preeclampsia.

Step 4. Preeclampsia induced either fetal death or live birth at the subsequent week (depicted in our DAG using a dashed arrow [Figure S1]). This one-week time frame was chosen for simplicity since timing of this outcome does not impact the simulated study's results. Preeclampsia increased the risk of fetal death (Methods S4).

Step 5. We determined final potential pregnancy outcomes by identifying the first pregnancy and preeclampsia outcome under each treatment. If the pregnancy outcome occurred first, then it was used. Otherwise, the pregnancy experienced preeclampsia and was assigned the preeclampsia-induced pregnancy outcome. "Observed" outcomes corresponded to assigned treatment and assumed perfect treatment persistence (Methods S5, Figure S3).

Step 6. We generated post-index prenatal care visits at gestational weeks 10-42. Prenatal encounters under mild hypertension mirrored clinical recommendations.[29] Moderate and severe hypertension increased weekly prenatal care probabilities (Methods S6). Rurality and treatment did not affect prenatal encounter frequency.

*Constructing Study Samples*

For each target population, we created 6 study samples distinguished by the target amount of missing pregnancy outcome data among non-initiators (low = 5% versus high = 20%)[5,30–33] and missing data mechanism (Figure 2). Three missing data mechanisms were explored:
1) 0% MAR and 100% MNAR – all missing data due to unobserved miscarriage (MNAR) and none due to measured factors (i.e., hypertension severity and rurality; MAR);

2) 50% MAR and 50% MNAR – an even split between missingness due to measured factors and missing due to unobserved miscarriage; and

3) 100% MAR and 0% MNAR – all missing data due to measured factors (Table S2).

These scenarios reflected Figures 1B and 1C where increased hypertension severity and rurality increased probability of missingness (Methods S7-S9, Table S3). This may reflect electronic health records data where severity increased the probability of being referred elsewhere for elevated care (i.e., censored) while rurality increased the probability that the person went elsewhere for care (i.e., censored). We further assumed that miscarriages may be unobserved, which was more likely earlier in gestation.

*Selecting Analytic Samples*

We created 3 analytic samples from each study sample (Figure 3).

1. *Observed pregnancies:* All pregnancies, regardless of missingness (i.e., "prenatal approach").

2. *Observed outcomes:* All pregnancies with observed outcomes (i.e., "outcome-based approach").

3. *Observed deliveries*: All pregnancies with an observed stillbirth (fetal death ≥20 weeks' gestation) or live birth (i.e., "outcome-based approach", limiting to deliveries).

*Analyzing the Analytic Samples*

Among the observed deliveries and observed outcomes analytic samples, risks were calculated by standardizing stratified risks to the joint severity and rurality distribution in the analytic sample.[35] For the observed pregnancies analytic samples, we estimated the risks using an Aalen-Johansen estimator within severity and rurality strata, modeling miscarriage and non-preeclamptic live births as competing events.[34,35] We accounted for confounding and informative censoring by standardizing stratified risks to the analytic samples' joint severity and rurality distribution.[36] We calculated RDs and RRs by contrasting the standardized risks (Methods S10).

True values were established using potential outcomes from the target population under full follow-up (i.e., no missing outcomes). We calculated bias in the risk, RD (per 100 pregnancies), and (log-transformed) RR by subtracting the true value for a parameter from the observed estimate.

Among observed pregnancies analytic samples, we calculated full bounds on the estimated risks, RDs, and RRs under four assumptions about censored pregnancies:[23,24]

1. All experienced prenatal preeclampsia,
2. All initiators experienced prenatal preeclampsia while non-initiators did not,
3. All non-initiators experienced prenatal preeclampsia while initiators did not, and
4. None experienced prenatal preeclampsia.

This simulation was conducted using R statistical software (R Foundation for Statistical Computing, Vienna, Austria). All code is available on GitHub ([LINK TO FOLLOW UNBLINDING]). ChatGPT was used for code de-bugging and editorial review of specific paragraphs (OpenAI, San Francisco, California).

**4. Simulation Findings**

True absolute risks of preeclampsia were 37.4% among non-initiators and ranged from 19.0% to 42.1% among initiators across the 6 target populations (Table 2). The RDs and RRs were furthest from the null when initiation increased the risk of miscarriage and decreased the risk of preeclampsia (RD=-18.4 per 100 pregnancies, RR=0.51) and were null when initiation affected neither miscarriage nor preeclampsia.

Bias in risks, RDs, and RRs differed across analytic samples and scenarios. This section focuses on scenarios where ~20% of pregnancies were missing outcomes and are reported in turn by the percentage of MNAR versus MAR pregnancies: only MNAR (i.e., 100% missing due to miscarriage), both MAR and MNAR (i.e., 50% missing due to severity/rurality and 50% due to miscarriage), and only MAR (i.e., 100% missing due to severity/rurality).

***Only MNAR (0% Severity/Rurality, 100% Miscarriage)***

Absolute risks were overestimated in all analytic samples (Figures S4-S5).

*Treatment Did Not Affect Preeclampsia.*

Estimated RDs and RRs were biased in scenarios where treatment increased or decreased the risk of miscarriage, though the magnitude and direction was similar across analytic samples (Figures 4A and 4D). For example, the RDs estimated from the three analytic samples ranged from -0.3 to 0.1 per 100 pregnancies (bias: 10.1 to 10.6 per 100 pregnancies) when treatment increased the risk of miscarriage (Tables S4-S6). Effect estimates were unbiased across analytic samples when treatment did not affect miscarriage.

*Treatment Decreased the Risk of Preeclampsia.*

Results were similar, though bias persisted in the RDs estimates from the three analytic samples when treatment initiation had no effect on miscarriage (Figures 5A and 5D). In that scenario, bias in the estimated RDs (per 100 pregnancies) ranged from -3.4 to -2.8 (Tables S4-S6).

**Both MAR and MNAR (50% Severity/Rurality, 50% Miscarriage)**

Absolute risks were overestimated among analytic samples when initiation decreased the risk of miscarriage. However, absolute risks were underestimated in some observed deliveries and observed outcomes analytic samples when treatment had no direct effect on preeclampsia (Figures S4-S5).

*Treatment Did Not Affect Preeclampsia.*

The observed deliveries analytic sample returned the most biased RD and RR estimates when treatment affected the risk of miscarriage (Figures 4B and 4E). For example, when initiation increased the risk of miscarriage, the estimated RD was -0.1 per 100 pregnancies (bias: 10.4) among the observed deliveries, -8.6 (bias: 1.9) among the observed outcomes, and -7.1 (bias: 3.4) among the observed pregnancies analytic samples (Tables S4-S6). RD and RR estimates were unbiased when treatment did not affect miscarriage.

Bias in the RD and RR estimates from the observed outcomes and observed pregnancies analytic samples decreased compared to scenarios where all missingness was due to unobserved miscarriage (MNAR) (Figures 4A and 4B, Tables S5-S6). For example, among the observed pregnancies analytic sample, the bias in the RDs (per 100 pregnancies) ranged from -5.5 to 10.2 when missingness was MNAR compared to -2.4 to 3.4 when missingness was due to both unobserved miscarriage and measured factors (MAR and MNAR).

*Treatment Decreased the Risk of Preeclampsia.*

Results were similar, though depended upon effect measure. Of the three analytic samples, the observed deliveries returned the most biased RR estimates across all scenarios. This was not true for the RD estimates when initiation increased the risk of miscarriage (Figures 5B and 5E).

RD and RR estimates among the observed pregnancies analytic sample were less biased compared to scenarios where all missingness was MNAR (Figures 5A and 5B). However, that did not hold for the RR from the observed outcomes analytic sample when treatment did not affect the risk of miscarriage: bias in the log-transformed RRs increased from 0.00 (RR=0.65) when all missingness was MNAR to -0.1 (RR=0.59) when both MAR and MNAR (Table S5).

***Only MAR (100% Severity, 0% Miscarriage)***

Absolute risks estimated from the observed deliveries analytic sample were higher than the true risks, while absolute risks estimated from the observed outcomes analytic sample were lower (Figures S4-S5). Absolute risk estimates from the observed pregnancies analytic samples were unbiased.

*Treatment Did Not Affect Preeclampsia.*

Effect estimates from the observed pregnancies analytic samples were unbiased. Estimates from the observed deliveries analytic sample were the most biased when initiation had a non-null effect on miscarriage (Figures 4C and 4F). RD estimates from the observed outcomes and observed pregnancies analytic samples were similar (bias range: -1.8 to 0.7 per 100 pregnancies and -0.3 to 0.3, respectively) (Tables S4-S6). However, RRs diverged when initiation increased the risk of miscarriage: they were 0.62 from the observed outcomes analytic sample and 0.71 from the observed pregnancies analytic sample, while the true value was 0.72 (bias: -0.15 and -0.01, respectively) (Tables S4-S6).

*Treatment Decreased the Risk of Preeclampsia.*

Results were similar, but RRs diverged even more (Figures 5C and 5F). For example, when initiation increased the risk of miscarriage, the estimated RRs were 0.38 among the observed outcomes analytic sample and 0.50 among the observed pregnancies analytic sample, while the truth was 0.51 (bias: -0.29 and -0.02, respectively) (Tables S5-S6). In fact, when initiation had no effect on the risk of miscarriage, the estimated RR from the observed outcomes analytic sample was equivalently biased as the RR estimated from the observed deliveries analytic sample (RR=0.55, bias: -0.17).

### *All Cohorts*

Full bounds included the absolute risks, RD, and RR estimates from the observed pregnancies and observed outcomes analytic samples (Figures S6-S7). There was less bias in estimated absolute risks, RDs, and RRs when ~5% versus ~20% of pregnancies were missing outcomes (Tables S4-S6, Figures S8-S11). Simulated pregnancies matched expected distributions for each scenario (Tables S8-S11).

### 5. Discussion

When all missingness was due to unobserved miscarriage (MNAR), estimated absolute risks, RDs, and RRs were biased regardless of pregnancy identification strategy used. RDs and RRs estimated from the analytic sample of observed pregnancies (using standardization to account for missing data) became less biased as the proportion of missingness due to measured factors (versus unmeasured factors) increased, which was not true for the other analytic samples. RDs and RRs estimated among observed deliveries were often the most biased. Only the observed pregnancies analytic sample consistently estimated the absolute risks without bias when all missingness was due to measured covariates. All analytic samples returned unbiased RD and RR estimates when treatment affected neither miscarriage nor preeclampsia.

Our findings align with previous studies demonstrating that conditioning on delivery can lead to biased effect estimates for prenatal exposures.[7,8,37–40] Moreover, our results corroborate the potential for selection bias induced by conditioning on post-exposure events, showing that restricting analyses to pregnancies with observed outcomes (i.e., "outcome-based approach") can yield biased effect estimates;[3,4] in fact, estimates may be just as biased as analyses restricted to deliveries. We also found that including pregnancies with missing outcomes (i.e., "prenatal approach") only resolved this issue to the extent that the causes of missing outcomes were measured and could be accounted for in the analysis. Bias similarly persisted when missingness was due to the true value of the outcome (i.e., miscarriage), as has been widely discussed in single outcome settings.[16,21] These findings underscore the importance of understanding the sources of missing data.

This article extends the sparse extant literature on bias due to missing outcomes for analyses with competing events. Current work has focused on missing covariates or missing the type of outcome (e.g., knowing death occurred but not the cause).[41–43] This differs from the situation faced by perinatal epidemiologists where a pregnancy's outcome may be entirely unknown. Our study's DAG assumed a multinomial outcome, which aligns with current definitions for potential outcomes in settings with competing events and allowed straightforward application of established rules for evaluating selection/missing data bias from DAGs in single outcome settings.[42,44–47] Empirical simulation findings conformed with theoretical expected results suggesting that this approach may be helpful in studies of prenatal exposures.

*Limitations*

Simulations necessarily simplify a complex problem to study a specific aspect of it, in this case bias due to pregnancy cohort creation. Additional considerations are required for real studies.[2] For example, generated data represented an idealized scenario in which gestational age estimates were perfect allowing accurate identification of the target population. Further, all simulated individuals initiated prenatal care at 9 weeks of gestation. This mirrors the average timing of prenatal care initiation in the U.S. but does not reflect the substantial variability around that timing.[48] This is not realistic for real-world populations but simplified analyses. Similarly, we assumed that treatment decisions were made and documented at prenatal encounters to avoid immortal time bias. Studies using real data must consider delays between time zero, treatment changes, and measurement.[49–52] Finally, we did not include a scenario where treatment increased the risk of the study outcome, as it is unlikely for this hypothetical study;[26] however, results would be comparable to scenarios where antihypertensive initiation decreased preeclampsia risk.

Due to these simplifying assumptions, these findings may have limited utility for predicting the magnitude and direction of bias in more complex scenarios. The amount and degree of bias in real analyses will depend upon the underlying data generating mechanisms. In the scenarios explored here, the RDs were less biased than the RRs; however, this is likely an artifact of our data generating approach as opposed to a finding that should be extrapolated to all studies. The most biased effect measure will depend upon the misestimation of risks, which is also critical for patient-provider decision-making.

Not all analytic issues were considered in the simulation. Severity or rurality might not be strong predictors of censoring in all data sources (e.g., claims), but they serve as examples of any measured variable (e.g., state-based abortion policies in the context of the U.S.). Further, the direction of their effects on observation of pregnancy outcomes may depend on the data source: for example, an academic medical system with maternal-fetal medicine specialists may be more likely to observe outcomes for pregnancies with severe hypertension but not their early prenatal encounters, potentially introducing immortal time if not addressed.[49] Investigators must carefully consider data- and population-specific sources of censoring. Further, we did not distinguish induced and spontaneous abortions. Instead, we assumed that no one intended to terminate their pregnancy–an inclusion criterion for the trial this work mirrored.[26] Real studies must consider if and how to distinguish these outcomes.[12] Finally, we did not consider situations in which the exposure window of interest begins at the last menstrual period, as the time between conception and first prenatal visit is arguably immortal,[49] which was outside the scope of this study.

***Implications for Analyses of Healthcare Data***

Bias in RD and RR estimates persisted among the observed deliveries and observed outcomes analytic samples when missingness was due to a measured variable (i.e., severity; MAR) but not among the all observed pregnancies analytic sample. Studies, then, that include all pregnancies where censoring is due to measured variables can utilize statistical approaches such as inverse probability of censoring weights to obtain unbiased estimates of the total effects of prenatal exposures.[17] While unlikely for all studies, this may be particularly important for studying exposures later in pregnancy (e.g., ≥13 weeks' gestation) where missing outcomes are more likely due to measurable factors instead of pregnancy loss or termination. However, these approaches are only available if investigators identify pregnancies with both observed and unobserved outcomes.

Including pregnancies with unobserved outcomes did not prevent bias when the outcome was missing due to miscarriage. This suggests that studies of early pregnancy exposures (e.g., <13 weeks' gestation)—where a higher proportion of pregnancies may be missing outcomes due to early loss or termination—must consider sensitivity analyses to grapple with the biases outlined in this paper, particularly when exposure has a non-null effect on the outcome or miscarriage, or there are unmeasured common causes of exposure and miscarriage. Sensitivity analyses exist to investigate this source of bias among cohorts that condition upon observed pregnancy outcomes.[19] However, identifying pregnancies with unobserved outcomes offers more approaches, as the extent of missingness is known. Re-analyzing the data under different assumptions about the outcomes experienced by censored pregnancies provided the full potential range of the true treatment effect estimate.[53] Investigators may also consider statistical methods designed to address MNAR outcome data under strict assumptions.[25,54,55] At the least, investigators must discuss and acknowledge these issues when interpreting study findings.

## 6. Conclusion

When unobserved outcomes were due to unobserved miscarriage, estimated risks and treatment effect estimates were similarly biased across pregnancy identification strategies. However, as the proportion of missingness due to measured factors increased, including pregnancies with unobserved outcomes resulted in less biased estimates. Although this pregnancy identification approach does not prevent bias from missing outcome data, it allows meaningful investigation of the potential impact of this bias on study findings.

# Tables

**Table 1.** Tabular description of three populations for understanding selection bias, realizations of the populations in a study of prenatal exposures on pregnancy outcomes using real-world data, and realizations in our simulation study.

| Population | Definition | Realization in Study of Prenatal Exposure on Pregnancy Outcomes Using Real-World Data | Realization in Simulated Data |
|---|---|---|---|
| Target Population | The population in which investigators hope to estimate the treatment effect without bias. | All pregnancies with a study-qualifying encounter and a non-zero probability of being assigned to both exposure groups. | All simulated pregnancies prior to inducing missing outcome data. |
| Study Sample | The population captured by the data source. The data captured reflects the realities of the data source (e.g., missing data). | All pregnancies in the data source that have ≥1 encounter that qualifies for study entry and a non-zero probability of being assigned to both exposure groups. | All simulated pregnancies after inducing missing outcome data (i.e., some pregnancies' outcomes are missing). |

| Analytic Sample | The population included in a study from the Study Sample. | All pregnancies identified by a pregnancy identification algorithm (i.e., "prenatal approach" or "outcome-based approach") with ≥1 encounter that qualifies for study entry. | 1. Observed Deliveries: simulated pregnancies with observed deliveries, 2. Observed Outcomes: simulated pregnancies with observed outcomes, and 3. Observed Pregnancies: all simulated pregnancies including those with missing outcomes |

**Table 2.** True risk, risk difference per 100 pregnancies (RD), and risk ratio (RR) estimates for the effect of antihypertensive initiation versus non-initiation on the risk of prenatal preeclampsia from 6 scenarios. Values were derived by using the potential outcomes under full follow-up among the 5,000,000 pregnancies randomized at baseline.

| Effect of Treatment on… | | Risk, Initiators | Risk, Non-Initiators | Risk Difference[a] | Risk Ratio |
| --- | --- | --- | --- | --- | --- |
| Miscarriage | Preeclampsia | | | | |
| Decrease | Decrease | 27.0% | 37.4% | -10.4 | 0.72 |
| No Effect | Decrease | 24.2% | 37.4% | -13.2 | 0.65 |
| Increase | Decrease | 19.0% | 37.4% | -18.4 | 0.51 |
| Decrease | No Effect | 42.1% | 37.4% | 4.7 | 1.13 |
| No Effect | No Effect | 37.5% | 37.4% | 0.1 | 1.00 |
| Increase | No Effect | 26.9% | 37.4% | -10.5 | 0.72 |

[a] Risk difference is presented per 100 pregnancies.

**Figures and Figure Legends**

**Figure 1.** Directed acyclic graphs for a study assessing the effect of a prenatal exposure E on multinomial study outcome D(j) with risk factor C and indicator for missing pregnancy outcome S. A box around S = 1 indicates that only individuals with an observed pregnancy outcome are included in the analytic cohort.

(A) 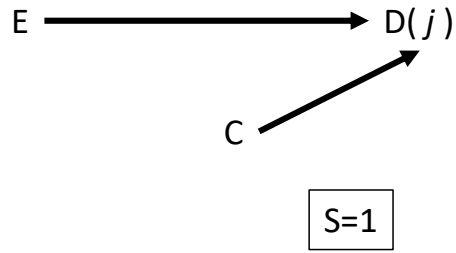 (B) 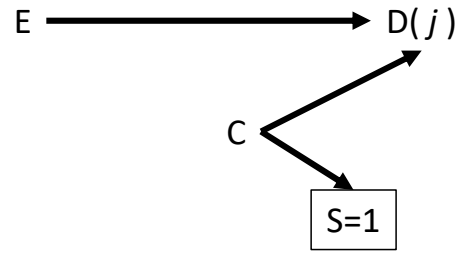 (C) 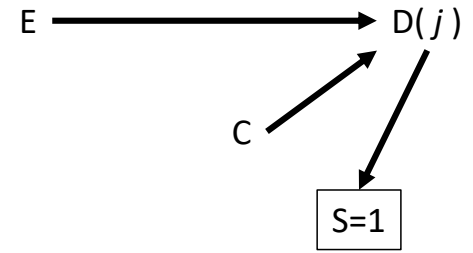

**Figure 2.** Depiction of scenarios distinguished according to the target percentage of non-initiators with unobserved pregnancy outcomes and causes of unobserved outcomes. Percentages for causes of unobserved outcomes correspond to the target percentage of unobserved outcomes attributable to that source.

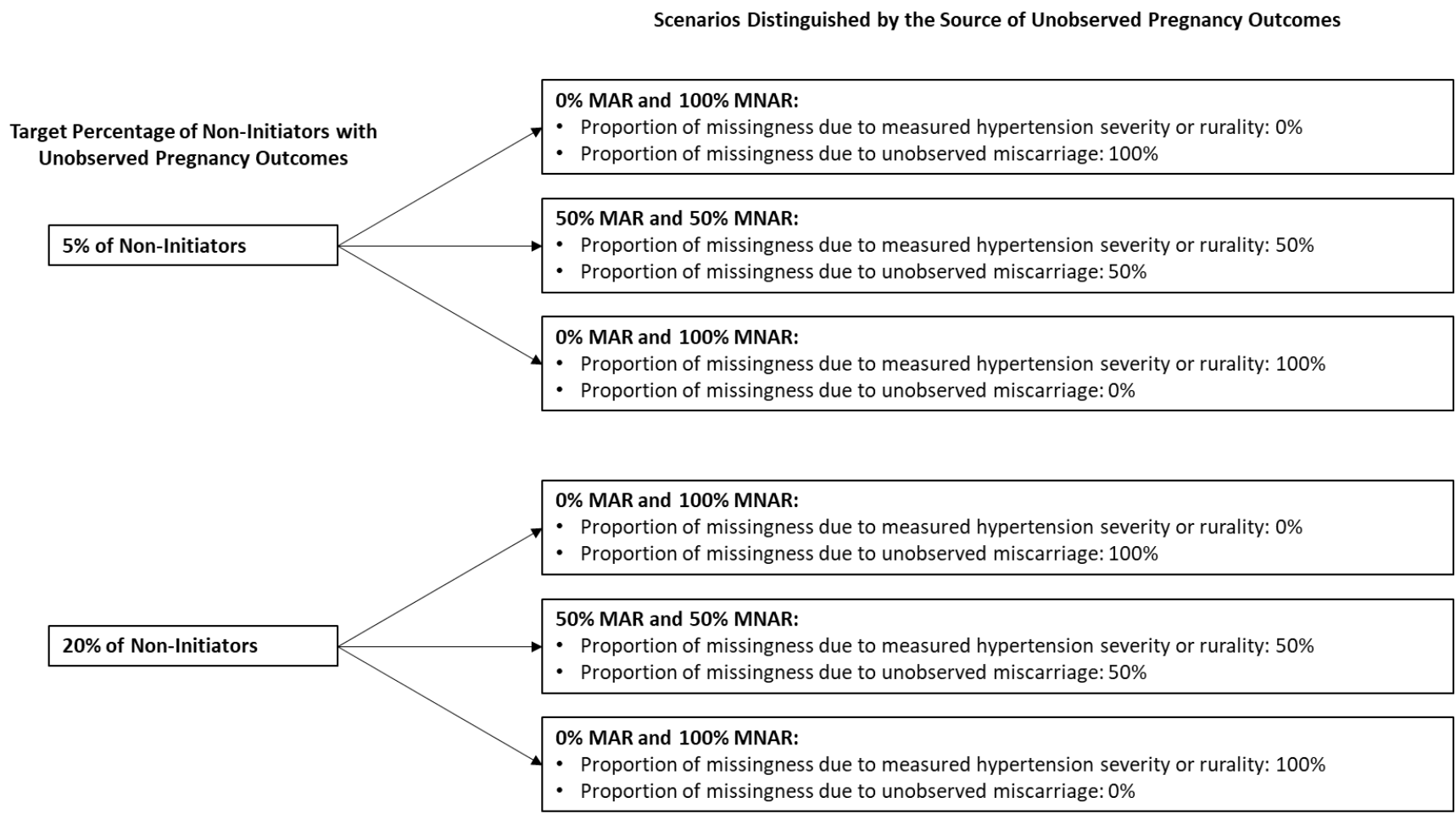

**Figure 3.** Diagram illustrating the relationships between the target population, study sample, and each analytic sample.

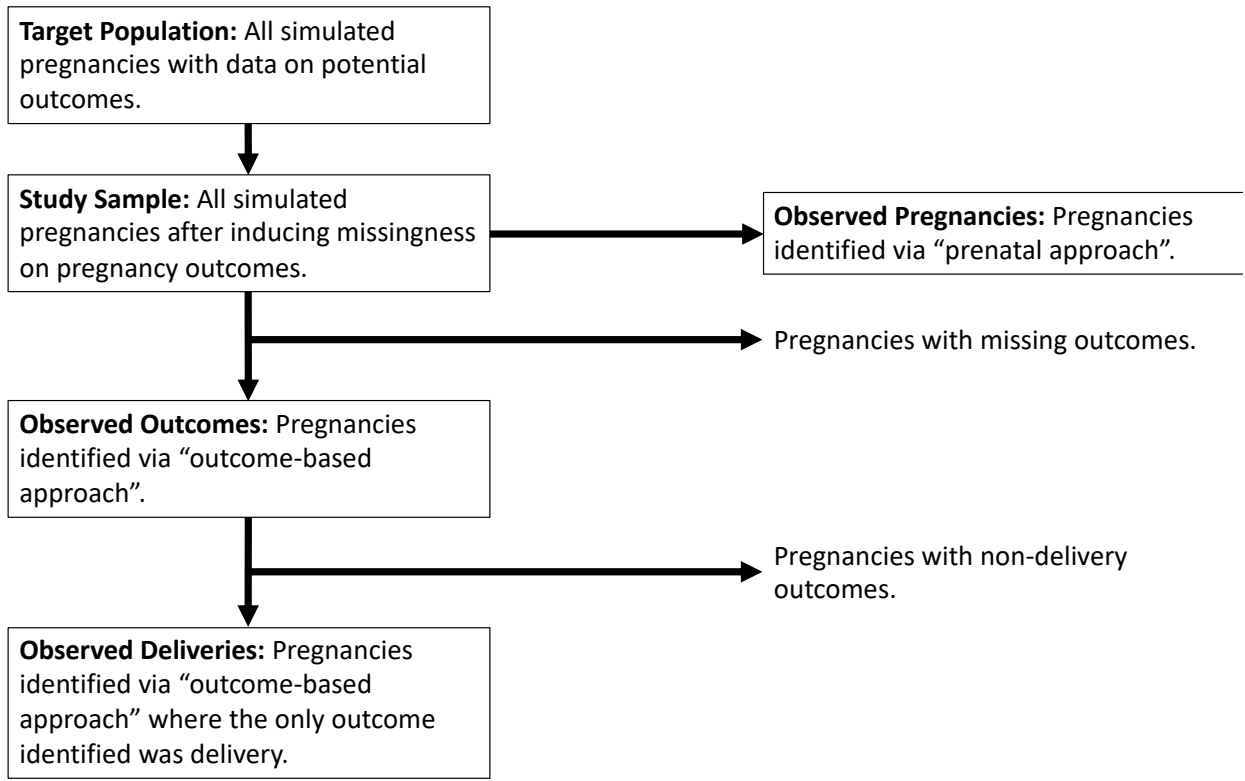

**Figure 4.** Results from scenarios where antihypertensive initiation had no direct effect on the risk of preeclampsia and ~20% of pregnancies were missing outcomes. The x-axis delineates scenarios according to the effect of initiation on miscarriage (decreased, no effect on, or increased risk). Results include bias in the risk difference (per 100 pregnancies) when pregnancy outcomes were (A) Missing Not At Random (MNAR), (B) a mix of Missing At Random (MAR) and MNAR, and (C) all MAR as well as bias in the risk ratio (log-transformed) when pregnancy outcomes were (D) MNAR, (E) a mix of MAR and MNAR, and (F) MAR. Bias was calculated by subtracting the true value for a parameter from the analytic sample's estimate. The horizontal, dashed line represents no bias. Analytic samples are differentiated by shape and color.

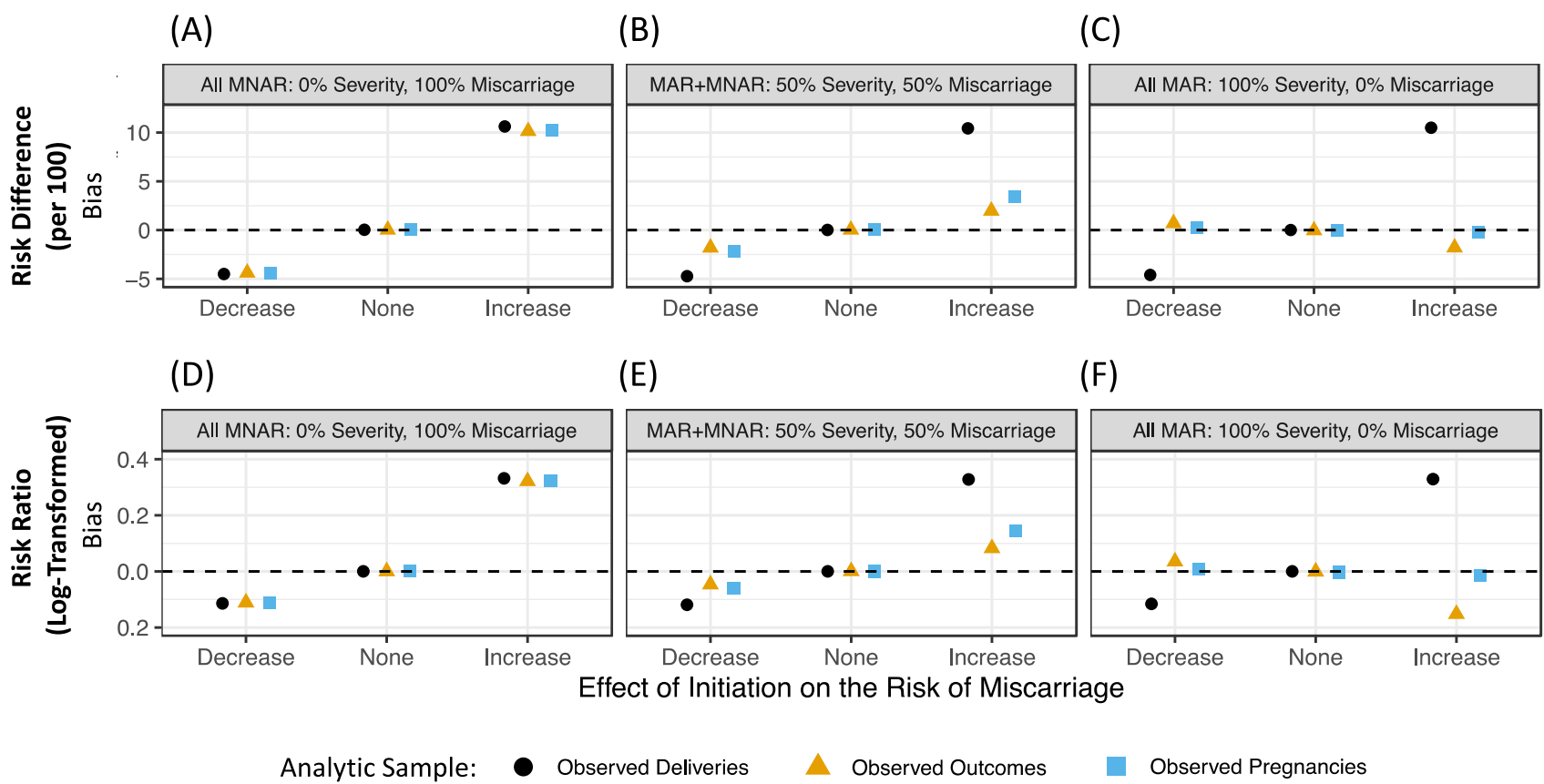

**Figure 5.** Analytic results from those scenarios where antihypertensive initiation decreased the risk of (prevented) preeclampsia and ~20% of pregnancies were missing outcomes. The x-axis delineates scenarios according to the effect of initiation on miscarriage (decreased, no effect on, or increased risk). Results include bias in the risk difference (per 100 pregnancies) when pregnancy outcomes were (A) Missing Not At Random (MNAR), (B) a mix of Missing At Random (MAR) and MNAR, and (C) all MAR as well as bias in the risk ratio (log-transformed) when pregnancy outcomes were (D) MNAR, (E) a mix of MAR and MNAR, and (D) MAR. Bias was calculated by subtracting the true value for a parameter from the analytic sample's estimate. The horizontal, dashed line represents no bias. Analytic samples are differentiated according to shape and color.

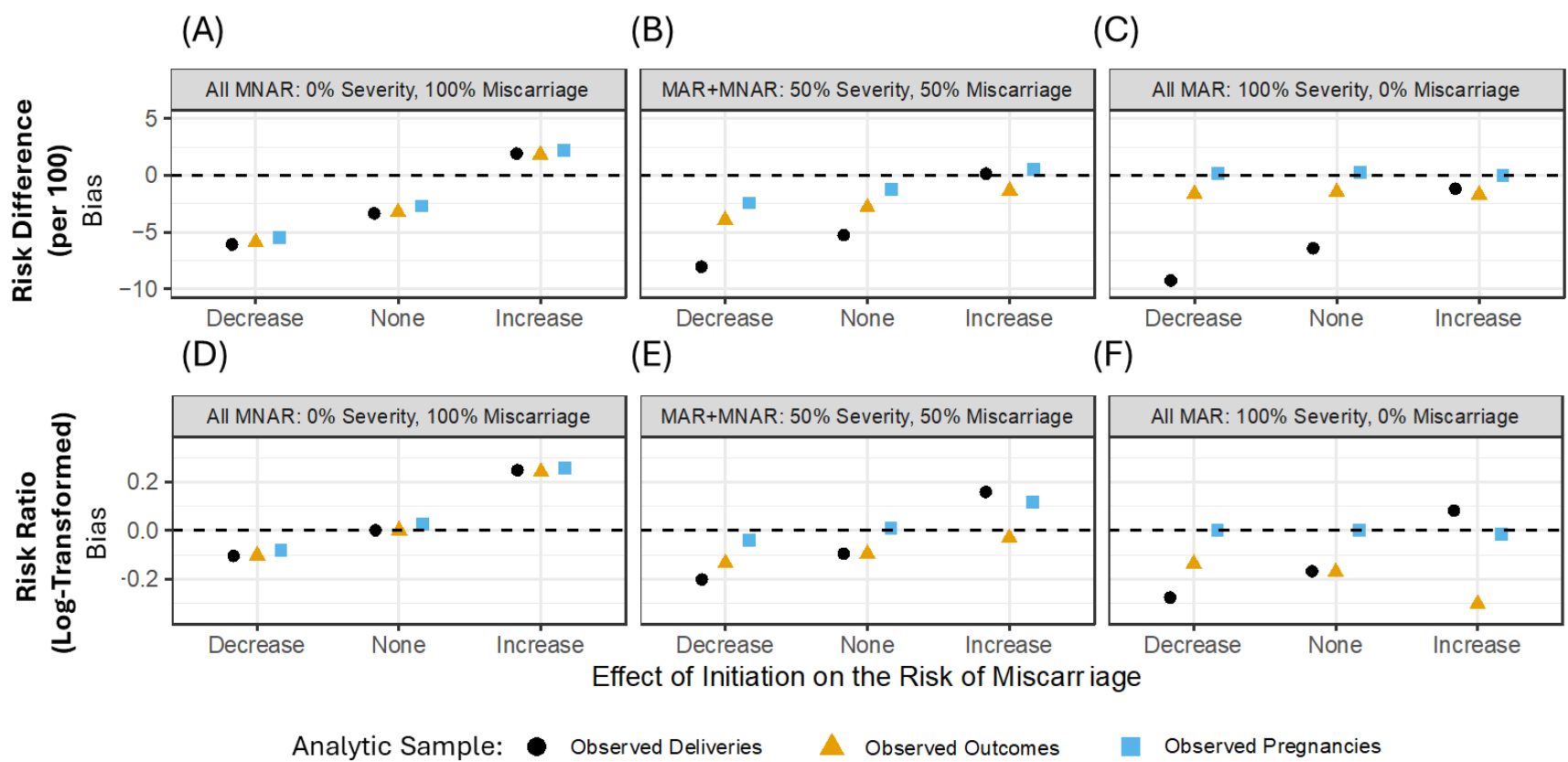

**Supplemental Document**

# Table of Contents

**Supplemental Methods**

**Methods S1.** Contextualizing our directed acyclic graph notation within other literature on competing events.

To our knowledge, there is not widespread agreement on how to represent competing events in directed acyclic graphs (DAG). One proposed approach involves drawing competing events as mediators.[1,2] Another involves drawing each potential value of the multinomial outcome going into one outcome node.[3] These approaches are particularly salient for thinking about direct, indirect, and separable effects in analyses with competing events.[4]

Our conceptualization differs from these proposed approaches by instead showing the outcome as a time-varying multinomial outcome. This aligns with proposals to analyze pregnancy as a multistate or multinomial outcome and with how we conceptualize potential outcomes for studies with competing events.[1,2,5–8] There are merits to both approaches, but we use this approach because it allows us to apply well-known rules for evaluating whether selection on an outcome may induce bias.[9–13]

**Methods S2.** Methodology for deriving weekly potential pregnancy outcomes.

Pregnancies were simulated in weeks from conception. As such, we index these methods from conception. The main manuscript indexes gestational weeks from last menstrual period (LMP), as that is how gestational age is typically indexed for clinical/real-world purposes. Throughout this manuscript, we assume that the LMP occurs 2 weeks prior to conception. All pregnancy outcomes were observed the week after they were determined.

We used a piecewise log-binomial regression to derive the weekly probability of miscarriage (spontaneous fetal death at ≤17 weeks from conception) based upon published estimates.[14–26] Increasing severity and rurality led to an increased risk of fetal death regardless of antihypertensive initiation. Initiation's effect on the risk of miscarriage depended upon scenario.

Probabilities were derived as:

$$p_{FD}(t) = \begin{cases} \exp(\beta_{0,t} + \beta_1 I_{moderate,0} + \beta_2 I_{high,0} + \beta_3 I_{low,1} + \beta_4 I_{moderate,1} + \beta_5 I_{high,1} + \beta_6 I_{rural}) & t < 16 \\ \exp(\beta_{0,t}) & t \geq 16 \end{cases}$$

where:
- $t$ = gestational week (from conception),
- $p_{FD}(t)$ = probability of fetal death at week $t$, observed week $t+1$
- $\beta_{0,t}$ = log-probability of fetal death at week $t$, observed week $t+1$, among non-initiators with low-severity disease residing in a non-rural area at baseline,
- $I_{low/moderate/high,0/1}$ = indicator for low/moderate/high severity hypertension and non-initiation/initiation (i.e., 0/1),
- $\beta_{1-5}$ = difference in the log-probability of fetal death among individuals with the corresponding severity and treatment level versus low-severity non-initiators,
- $I_{rural}$ = indicator for residing in a non-rural (0) or rural (1) area, and
- $\beta_6$ = difference in the log-probability of fetal death among individuals living in rural versus non-rural areas.

For potential outcomes under antihypertensive initiation and non-initiation, two pregnancy outcomes were possible for weeks 7-20 from conception (9-22 from LMP): continuing pregnancy or fetal death. Outcomes were selected via a Bernoulli random variable (i.e., $I_{FD}(t) \sim Bern(p_{FD}(t))$ where $I_{FD}(t)$ is an indicator for fetal death at week $t$, observed at week $t+1$).

Possible pregnancy outcomes after week 20 from conception (22 from LMP) included continuing pregnancy, live birth, and fetal death. As such, weekly pregnancy outcomes after week 20 were selected via a categorical random variable parameterized by the probability of continuing pregnancy ($p_{CONT}(t)$), the probability of fetal death ($p_{FD}(t)$), and the probability of live birth ($p_{LB}(t)$) (i.e., $I_{OUTCOME}(t) \sim Cat(K = 3, \mathbf{p} = (p_{CONT}(t), p_{FD}(t), p_{LB}(t)))$, where $I_{OUTCOME}(t)$ is an indicator for the pregnancy outcome at week $t$, observed at week $t+1$). These probabilities summed to 1 at each week. The probability of live birth at each gestational week was the same regardless of severity, rurality, and treatment. As such, the probability of continuing pregnancy at week $t$ was calculated by subtracting $p_{FD}(t)$ and $p_{LB}(t)$ from 1. At week 40 from conception, the probability of continuing pregnancy was 0.

Across all scenarios, $\beta_1 = \ln(2)$ (i.e., RR = 2), $\beta_2 = \ln(3)$ (i.e., RR = 3), and $\beta_6 = \ln(1.5)$ (i.e., RR = 1.5). The values of $\beta_3, \beta_4$, and $\beta_5$ depended upon the scenario, shown below.

| Effect of Initiation on the Risk of Miscarriage | $\beta$ values |
|---|---|
| Decreased | $\beta_3 = \ln(0.1), \beta_4 = \ln(0.5), \beta_5 = \ln(0.8)$ |
| No Effect | $\beta_3 = \beta_4 = \beta_5 = 0$ |
| Increased | $\beta_3 = \ln(5), \beta_4 = \ln(2), \beta_5 = \ln(1.1)$ |

**Methods S3.** Methodology for deriving weekly potential preeclampsia outcomes.

Pregnancies were simulated in weeks from conception. As such, we index these methods from conception. The main manuscript indexes gestational weeks from last menstrual period (LMP), as that is how gestational age is typically indexed for clinical/real-world purposes. Throughout this manuscript, we assume that the LMP occurs 2 weeks prior to conception. All pregnancy outcomes were observed the week after they were determined.

Probabilities of preeclampsia among non-initiators with low severity disease mirrored published estimates.[19,20,27,28] Increasing severity led to an increased risk of preeclampsia regardless of antihypertensive initiation, as did residing in a rural versus non-rural area. Initiation could modify the risk of preeclampsia. The probability of preeclampsia prior to week 17 from conception was 0. We used logistic regression to derive the weekly probability of preeclampsia for gestational weeks 17-40 from conception (observed at weeks 18-41):

$$p_{PE}(t) = expit(\gamma_{0,t} + \gamma_1 I_{moderate,0} + \gamma_2 I_{high,0} + \gamma_3 I_{low,1} + \gamma_4 I_{moderate,1} + \gamma_5 I_{high,1} + \gamma_6 I_{rural})$$

where
- $t$ = gestational week from conception;
- $p_{PE}(t)$ = probability of preeclampsia at gestational week $t$, observed at week $t + 1$;
- $\gamma_{0,t}$ = log-probability of preeclampsia at week $t$ among non-initiators with low-severity disease residing in a non-rural area;
- $I_{low/moderate/high,0/1}$ = indicator for low/moderate/high severity hypertension and non-initiation/initiation;
- $\gamma_{1-5}$ = difference in the log-odds of preeclampsia among individuals with the corresponding severity and treatment level versus low-severity non-initiators;
- $\gamma_6$ = difference in the log-odds of preeclampsia among individuals residing in a rural versus non-rural area at baseline; and
- $I_{rural}$ = indicator that the person resided in a non-rural (0) or rural (1) area.

Across all scenarios, $\gamma_1 = \ln(1.5)$ (i.e., OR = 1.5), $\gamma_2 = \ln(2)$ (i.e., OR = 2), and $\gamma_6 = \ln(2)$ (i.e., OR=2). The values of $\gamma_3, \gamma_4$, and $\gamma_5$ depended upon the scenario, as shown below.

| Effect of Initiation on the Risk of Miscarriage | $\beta$ values |
|---|---|
| Decreased | $\gamma_3 = \ln(0.2), \gamma_4 = \ln(0.5), \gamma_5 = \ln(0.8)$ |
| No Effect | $\gamma_3 = \gamma_4 = \gamma_5 = 0$ |

We used a Bernoulli random variable to determine the occurrence of preeclampsia at each week, observed at the next week (i.e., $I_{PE}(t) \sim Bern(p_{PE}(t))$ where $I_{PE}(t)$ is an indicator of preeclampsia at week $t$, observed at week $t + 1$).

**Methods S4.** Methodology used to derive weekly preeclampsia-induced pregnancy outcomes.

Pregnancies were simulated in weeks from conception. As such, we index these methods from conception. The main manuscript indexes gestational weeks from last menstrual period (LMP), as that is how gestational age is typically indexed for clinical/real-world purposes. Throughout this manuscript, we assume that the LMP occurs 2 weeks prior to conception. All pregnancy outcomes were observed the week after they were determined.

Weekly probabilities of preeclampsia-induced fetal death versus live birth ($p_{FD|PE}(t)$) were determined using a log-binomial model:

$$p_{FD|PE}(t) = \exp(\ln(p_{FD}(t)) + \alpha_0)$$

where
- $t =$ gestational week from conception;
- $p_{FD|PE}(t) =$ probability of preeclampsia-induced fetal death at week $t$, observed at week $t + 1$;
- $p_{FD}(t) =$ probability of fetal death at week $t$, observed week $t + 1$; and
- $\alpha_0 =$ difference in log-probability preeclampsia-induced fetal death after developing preeclampsia.

Throughout all simulations, $\alpha_0 = \ln(2.5)$. Severity did not impact the risk of preeclampsia-induced fetal death. We used a Bernoulli random variable to determine the occurrence of preeclampsia-induced fetal death versus live birth at weeks 17-41 (i.e., $I_{FD|PE}(t) \sim Bern(p_{FD|PE}(t))$ where $I_{FD|PE}(t)$ is an indicator of preeclampsia-induced fetal death at week $t$, observed at week $t + 1$).

**Methods S5.** Detailed process for selecting simulated pregnancies' observed outcomes from all generated potential outcome values.

We selected observed outcomes from the potential outcomes according to assigned treatment values using a stepwise process.

1. We assumed perfect treatment persistence: pregnancies that did not initiate antihypertensives at their indexing prenatal encounter were never treated, and pregnancies that initiated treatment were always treated at and after that encounter.
2. We considered those pregnancy and preeclampsia outcomes that corresponded to observed treatment.
3. To determine a pregnancy's outcome, we identified the first pregnancy outcome that was not "continuing pregnancy" and the first preeclampsia event.
    a. If the pregnancy outcome occurred first, we recorded that pregnancy outcome, which was observed at the subsequent week.
    b. If the first preeclampsia event occurred the same week as or prior to the first pregnancy outcome, then we recorded the prenatal preeclampsia outcome (observed the subsequent week). Further, we recorded the pregnancy outcome as the preeclampsia-induced pregnancy outcome at that gestational week, as opposed to the original pregnancy outcome at that week. The pregnancy outcome was also observed at the subsequent week, though we assumed that the prenatal preeclampsia occurred first.

**Methods S6.** Methodology for deriving weekly prenatal care encounters.

Prenatal encounters were simulated at weeks from conception. As such, we index these methods from conception. The main manuscript indexes gestational weeks from last menstrual period (LMP), as that is how gestational age is typically indexed for clinical/real-world purposes. Throughout this manuscript, we assume that the LMP occurs 2 weeks prior to conception. All pregnancy outcomes were observed the week after they were determined.

All pregnancies had a probability of 1 for a prenatal encounter at gestational week 7 from conception (9 from LMP). We then used a log-binomial model to determine probabilities for having a prenatal care encounter at week $t > 7$:

$$p_{PNC}(t) = \exp(\lambda_{0,t} + \lambda_1 I_{moderate} + \lambda_2 I_{high})$$

where
- $t$ = gestational week from conception
- $p_{PNC}(t)$ = probability of a prenatal care encounter at gestational week $t$,
- $\lambda_{0,t}$ = log-probability of a prenatal care encounter at gestational week $t$ for low severity hypertension,
- $I_{moderate/high}$ = indicator for moderate/high severity hypertension,
- $\lambda_1$ = difference in log-probability of prenatal care encounter for moderate versus low severity hypertension, and
- $\lambda_2$ = difference in log-probability of prenatal care encounter for moderate versus low severity hypertension.

We used a Bernoulli random variable to determine if a prenatal care encounter occurred (i.e., $I_{PNC}(t) \sim Bern(p_{PNC}(t))$ where $I_{PNC}(t)$ is an indicator for a prenatal care encounter at week $t$).

**Methods S7.** Identifying study samples distinguished by the amount of missing pregnancy outcome data and missingness mechanism.

Pregnancies were simulated in weeks from conception. As such, we index these methods from conception. The main manuscript indexes gestational weeks from last menstrual period (LMP), as that is how gestational age is typically indexed for clinical/real-world purposes. Throughout this manuscript, we assume that the LMP occurs 2 weeks prior to conception. All pregnancy outcomes were observed the week after they were determined.

We generated weekly probabilities of missing due to measured variables (severity or rurality) at week $t$ ($p_{miss,measured}(t)$) using a logistic regression with a balancing intercept term parameterized such that increasing severity and residing in a rural area were associated with a higher probability of missing (Methods S8, Table S3).[29,30] We used a Bernoulli random variable to determine if a pregnancy was missing due to measured variables (i.e., $I_{miss,measured}(t) \sim Bern(p_{miss,measured}(t))$ where $I_{miss,measured}(t)$ is an indicator for being missing due to measured variables at week $t$).

Pregnancies that ended in miscarriage could be missing outcomes due to miscarriage. We generated a probability of missingness due to miscarriage ($p_{miss,miscarriage}$) via logistic regression with a balancing intercept term (Methods S9, Table S3). The logistic regression was parameterized such that increasing gestational age at miscarriage was associated with a lower probability of missing. We used a Bernoulli random variable to determine if the miscarriage was missing (i.e., $I_{miss,miscarriage} \sim Bern(p_{miss,miscarriage})$ where $I_{miss,miscarriage}$ is an indicator for missing due to miscarriage.).

Final missingness was determined by identifying the first instance of missingness due to measured variables. If before the pregnancy outcome, the pregnancy was missing due to measured variables (severity or rurality) and censored at the most recent prenatal encounter. Pregnancies missing due to miscarriage were censored at their most recent prenatal encounter before the miscarriage.

**Methods S8.** Logistic regression model for weekly probabilities of missingness due to severity and rurality.

Pregnancies were simulated in weeks from conception. As such, we index these methods from conception. The main manuscript indexes gestational weeks from last menstrual period (LMP), as that is how gestational age is typically indexed for clinical/real-world purposes. Throughout this manuscript, we assume that the LMP occurs 2 weeks prior to conception. All pregnancy outcomes were observed the week after they were determined.

We generated weekly probabilities of being missing due to measured variables (severity and rurality) ($p_{miss,measured}(t)$) using a logistic regression with a balancing intercept term.[29,30] We specified a desired marginal probability of missingness over 34 weeks of follow-up (i.e., weeks 8 to 41 from conception, inclusive) ($p_{32,miss,measured}$) and then determined the corresponding weekly probability to achieve that marginal probability: $p_{34,miss,measured}(t) = 1 - (1 - p_{marginal,miss,measured})^{\frac{1}{34}}$ (Table S3). Weekly probabilities of missingness due to severity or rurality at weeks 8-41 from conception were determined via a logistic regression:

$$p_{miss,measured}(t) = expit\left(-\log\left(\frac{1}{p_{marginal,miss,measured}(t)} - 1\right) - \alpha_1 E(s) - \alpha_2 E(r) + \alpha_1 s + \alpha_2 r\right)$$

where
- $t$ = gestational week from conception;
- $p_{miss,measured}(t)$ = probability of missing due to the measured variables severity or rurality at week $t$;
- $s$ = categorical indicator for severity (0=low, 1=moderate, 2=severity);
- $r$ = indicator for residing in a rural area (0=non-rural, 1=rural);
- $E(s)$ = average severity level among non-initiators in the study sample;
- $E(r)$ = proportion of non-initiators in the study sample residing in a rural area at baseline (30%);
- $\alpha_1$ = difference in the log-odds of missing due to severity for moderate versus low and high versus moderate severity hypertension; and
- $\alpha_2$ = difference in the log-odds of missing for individuals residing in rural versus non-rural areas at baseline.

**Methods S9.** Logistic regression model used to derive the probabilities of missingness due to miscarriage.

The probability of missingness due to miscarriage was modeled via logistic regression with a balancing intercept term.[29,30] We specified a desired marginal probability of missingness due to miscarriage ($p_{marginal,miss,miscarriage}$) (Table S3) and incorporated this a the logistic regression:

$$p_{miss,miscarriage} = expit\left(-\log\left(\frac{1}{p_{marginal,miss,miscarriage}} - 1\right) - \delta_1 E(g) + \delta_1 g\right)$$

where
- $p_{miss,miscarriage}$ = probability of being missing due to miscarriage,
- $E(g)$ = average gestational age at miscarriage among pregnancies in the study sample that ended in miscarriage,
- $g$ = gestational age at miscarriage, and
- $\delta_1$ = difference in the log-odds of missing due to miscarriage for miscarriages that ended at gestational week $g$ versus $g - 1$.

**Methods S10.** Analyses conducted within each analytic sample.

*"Truth" for Each Analytic Sample*

We established the true values of study parameters within each analytic sample using each pregnancy's potential outcomes under full follow-up (i.e., no missingness) from the corresponding target population. Parameters were calculated as follows:
- Absolute risks: Number of potential outcomes with the event divided by the number of pregnancies in the target population,
- Risk difference (RD): (risk if all had initiated) – (risk if all had not initiated), and
- Risk ratio (RR): (risk if all had initiated) / (risk if all had not initiated).

*Observed Deliveries & Observed Outcomes Analytic Samples*

We used non-parametric direct standardization to account for confounding by hypertension severity and missingness due to rurality.[31] Specifically, we calculated the risks within each hypertension severity, rurality, and treatment stratum using simple proportions (i.e., number of events divided by the number of pregnancies at baseline) and then standardized each treatment's stratified risks to the joint severity and rurality distribution in the full analytic sample (i.e., population average treatment effect [ATE]). We then contrasted them using the same two parameters: RD and RR.

*Observed Pregnancies Analytic Sample*

We accounted for confounding and informative right-censoring using non-parametric direct standardization. (This approach worked appropriately for right-censoring because the censoring process due to severity was random within strata of severity.) We thus used an Aalen-Johansen estimator (treating non-preeclamptic miscarriage and livebirth as separate competing events) to estimate the risk of prenatal preeclampsia among observed pregnancies stratified according to hypertension severity, rurality, and treatment.[32] We then standardized these risks according to the joint severity and rurality distribution in the analytic sample. For those pregnancies with only their initial prenatal encounter observed, their time-to-event was recorded as a small constant (0.0001).

We further established the full set of bounds on the parameters via sensitivity analyses making four different assumptions about pregnancies with missing outcomes, as detailed in the manuscript.[33] Risks were estimated using simple proportions within strata of hypertension severity, rurality, and treatment and standardized to the joint severity and rurality distribution in the full analytic sample. We then calculated RDs and RRs using these standardized risk estimates.

**Supplemental Figures**

**Figure S1.** Directed acyclic graphs used to generate pregnancy and preeclampsia outcomes in the target populations. The variable t indexes the gestational week from a pregnancy's conception. Pregnancy and preeclampsia outcomes were generated at weeks 7-40 from <u>conception</u> and observed the week after they were generated (i.e., 8-41). The possible pregnancy outcomes were fetal death or continuing pregnancy for weeks 7-20; fetal death, live birth, or continuing pregnancy for weeks 21-39; and fetal death or live birth for week 40. Preeclampsia was possible at weeks 17-40 (observed at weeks 18-41). The dashed arrow indicates a deterministic outcome. Specifically, if a person has preeclampsia, they will have a preeclampsia-induced pregnancy outcome observed the next week (the same week the prenatal preeclampsia is observed but assumed to occur after the preeclampsia).

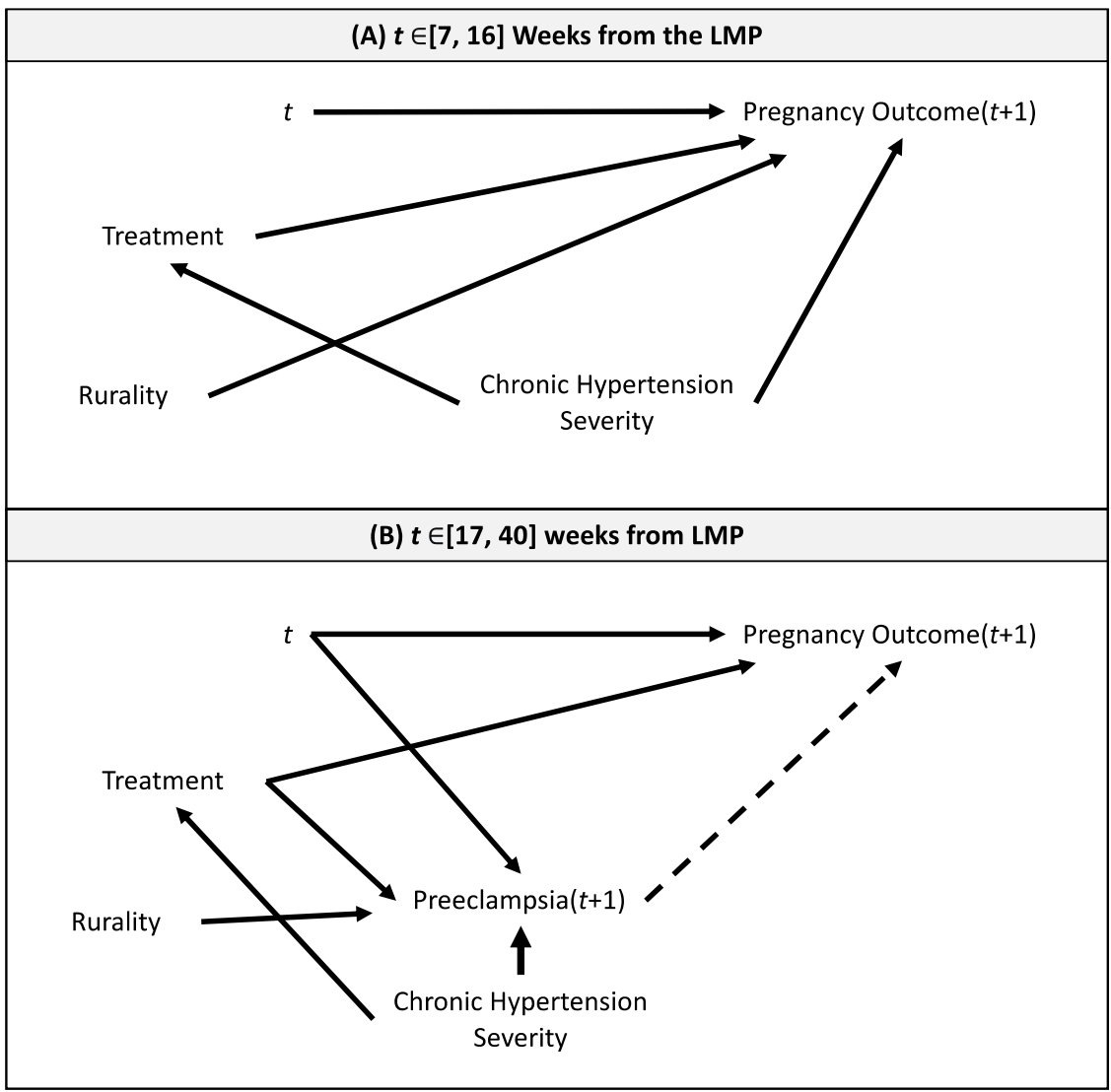

Note: Pregnancies were simulated in weeks from conception. As such, we index these methods from conception. The main manuscript indexes gestational weeks from last menstrual period (LMP), as that is how gestational age is typically indexed for clinical/real-world purposes. Throughout this manuscript, we assume that the LMP occurs 2 weeks prior to conception. All pregnancy outcomes were observed the week after they were determined.

**Figure S2.** Depiction of the data generation process for one pregnancy. Within tables, numbered columns represent the gestational week according to the number of weeks from conception (2 weeks after the last menstrual period).

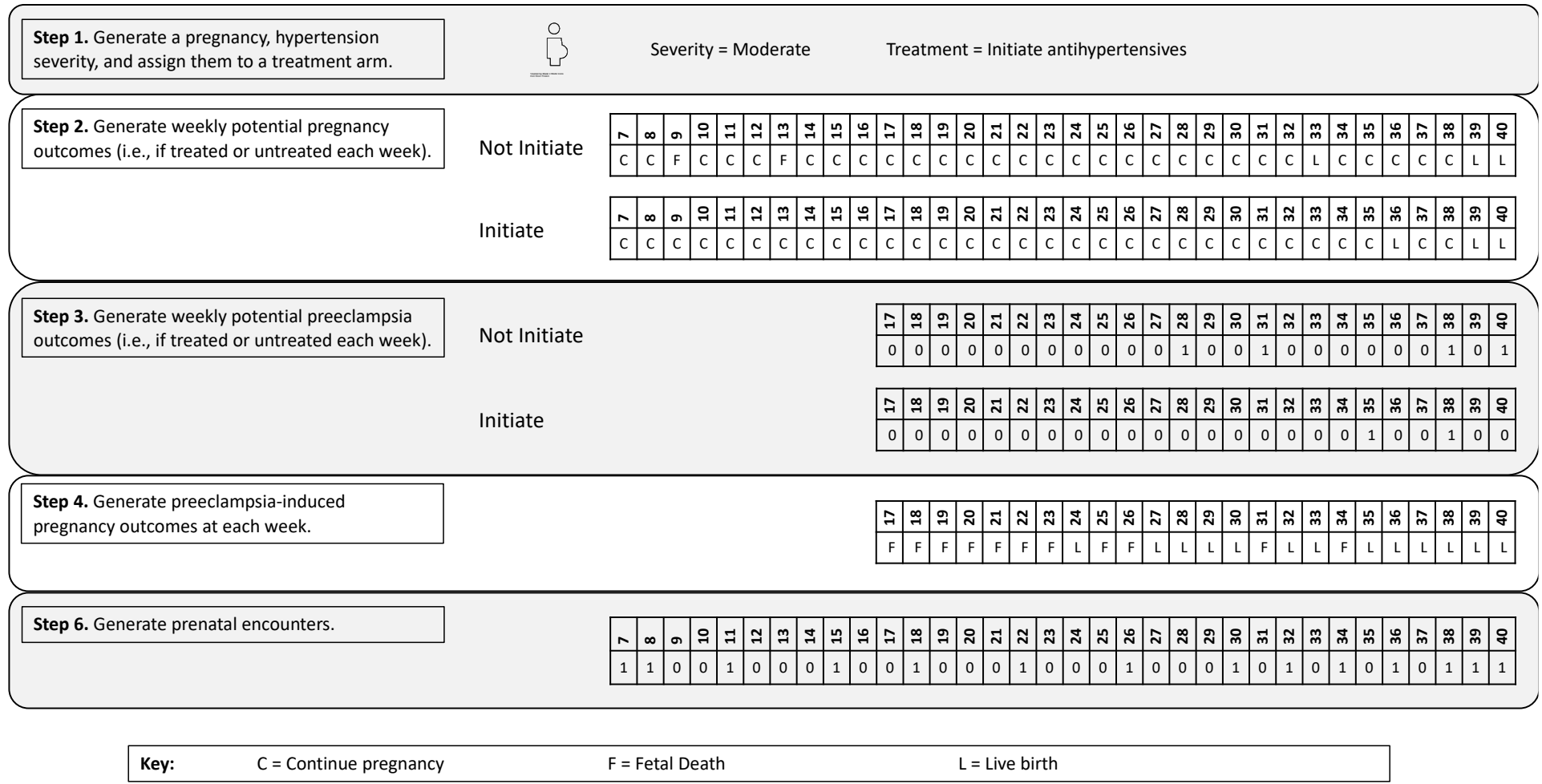

**Figure S3.** Figure depicting the data selection process from Step 5 of data generation for the pregnancy in Figure S2. Within tables, numbered columns represent the gestational week according to the number of weeks from conception (2 weeks after the last menstrual period).

Gather pregnancy and preeclampsia outcomes corresponding to the assigned treatment arm (initiation in Figure S1). Identify first pregnancy and preeclampsia outcomes.

Pregnancy outcomes:

| 7 | 8 | 9 | 10 | 11 | 12 | 13 | 14 | 15 | 16 | 17 | 18 | 19 | 20 | 21 | 22 | 23 | 24 | 25 | 26 | 27 | 28 | 29 | 30 | 31 | 32 | 33 | 34 | 35 | 36 | 37 | 38 | 39 | 40 |
|---|---|---|---|---|---|---|---|---|---|---|---|---|---|---|---|---|---|---|---|---|---|---|---|---|---|---|---|---|---|---|---|---|---|
| C | C | C | C | C | C | C | C | C | C | C | C | C | C | C | C | C | C | C | C | C | C | C | C | C | C | C | C | C | **L** | C | C | L | L |

Preeclampsia outcomes:

| 17 | 18 | 19 | 20 | 21 | 22 | 23 | 24 | 25 | 26 | 27 | 28 | 29 | 30 | 31 | 32 | 33 | 34 | 35 | 36 | 37 | 38 | 39 | 40 |
|---|---|---|---|---|---|---|---|---|---|---|---|---|---|---|---|---|---|---|---|---|---|---|---|
| 0 | 0 | 0 | 0 | 0 | 0 | 0 | 0 | 0 | 0 | 0 | 0 | 0 | 0 | 0 | 0 | 0 | 0 | **1** | 0 | 0 | 1 | 0 | 1 |

Because preeclampsia occurred first, determine final pregnancy outcome according to the pregnancy outcomes after preeclampsia.

Preeclampsia-induced pregnancy outcomes:

| 17 | 18 | 19 | 20 | 21 | 22 | 23 | 24 | 25 | 26 | 27 | 28 | 29 | 30 | 31 | 32 | 33 | 34 | 35 | 36 | 37 | 38 | 39 | 40 |
|---|---|---|---|---|---|---|---|---|---|---|---|---|---|---|---|---|---|---|---|---|---|---|---|
| F | F | F | F | F | F | F | L | F | F | F | L | L | L | L | F | L | L | F | **L** | L | L | L | L |

Outline final pregnancy timeline based upon the earlier outcomes. Pregnancy and preeclampsia outcomes are observed the week *after* they are observed.

| | 7 | 8 | 9 | 10 | 11 | 12 | 13 | 14 | 15 | 16 | 17 | 18 | 19 | 20 | 21 | 22 | 23 | 24 | 25 | 26 | 27 | 28 | 29 | 30 | 31 | 32 | 33 | 34 | 35 | 36 | 37 | 38 | 39 | 40 | 41 |
|---|---|---|---|---|---|---|---|---|---|---|---|---|---|---|---|---|---|---|---|---|---|---|---|---|---|---|---|---|---|---|---|---|---|---|---|
| Treatment values | 1 | 1 | 1 | 1 | 1 | 1 | 1 | 1 | 1 | 1 | 1 | 1 | 1 | 1 | 1 | 1 | 1 | 1 | 1 | 1 | 1 | 1 | 1 | 1 | 1 | 1 | 1 | 1 | 1 | 1 | 1 | 1 | 1 | 1 | |
| Observed preeclampsia indicator | | | | | | | | | | | | | | | | | | | | | | | | | | | | | | 1 | | | | | |
| Observed pregnancy outcome | | | | | | | | | | | | | | | | | | | | | | | | | | | | | | L | | | | | |

**Key:**  C = Continue pregnancy   F = Fetal Death   L = Live birth

**Figure S4.** Absolute risks from scenarios where antihypertensive initiation had no direct effect on the risk of preeclampsia and ~20% of pregnancies were missing outcomes. Results include bias (in percentage points) in the risk among (A) initiators and (B) non-initiators when outcomes were Missing Not At Random (MNAR), a mixture of MNAR and Missing at Random (MAR), or MAR. The x-axis delineates scenarios according to the effect of initiation on miscarriage (decreased, no effect on, or increased risk). The horizontal, dashed line represents a no residual bias. Analytic samples are differentiated by shape and color.

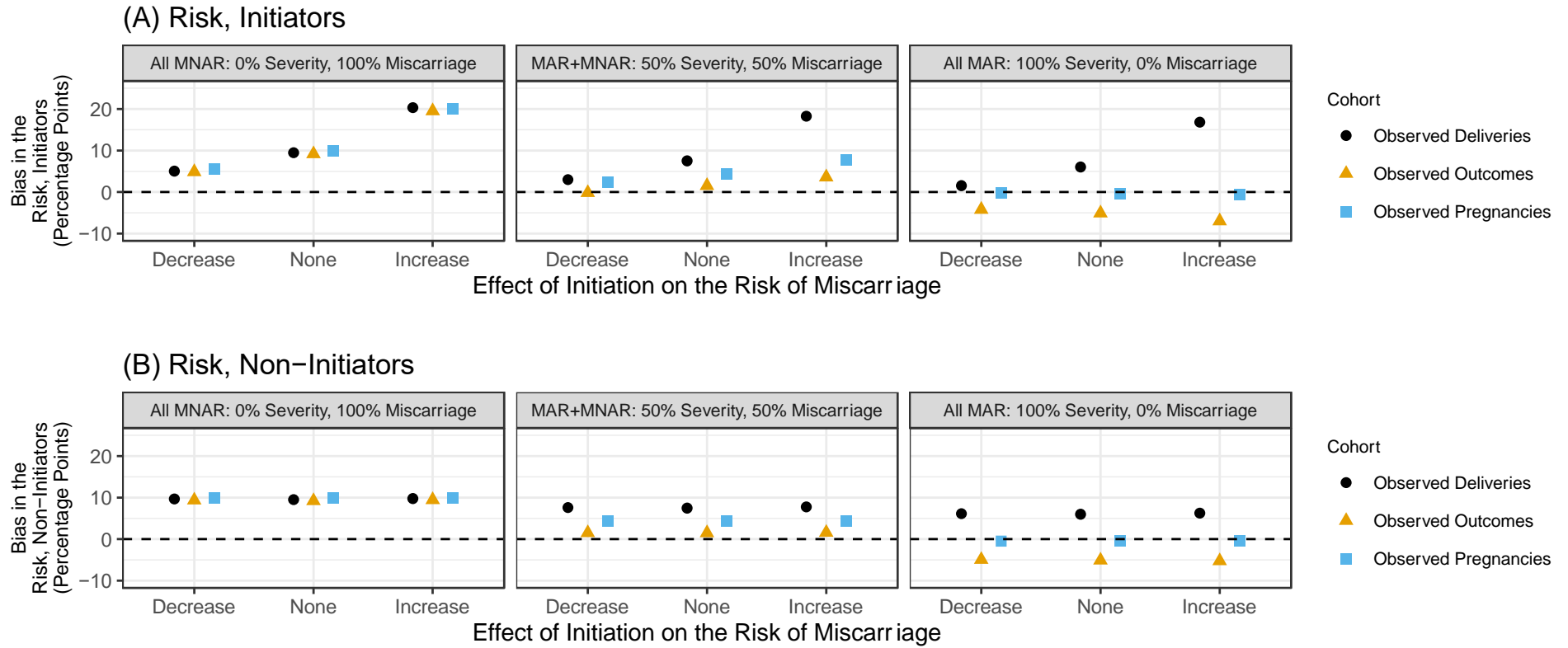

**Figure S5.** Absolute risks from scenarios where antihypertensive initiation decreased the risk of preeclampsia and ~20% of pregnancies were missing outcomes. Results include bias (in percentage points) in the risk among (A) initiators and (B) non-initiators when outcomes were Missing Not At Random (MNAR), a mixture of MNAR and Missing at Random (MAR), or MAR. The x-axis delineates scenarios according to the effect of initiation on miscarriage (decreased, no effect on, or increased risk). The horizontal, dashed line represents a no residual bias. Analytic samples are differentiated by shape and color.

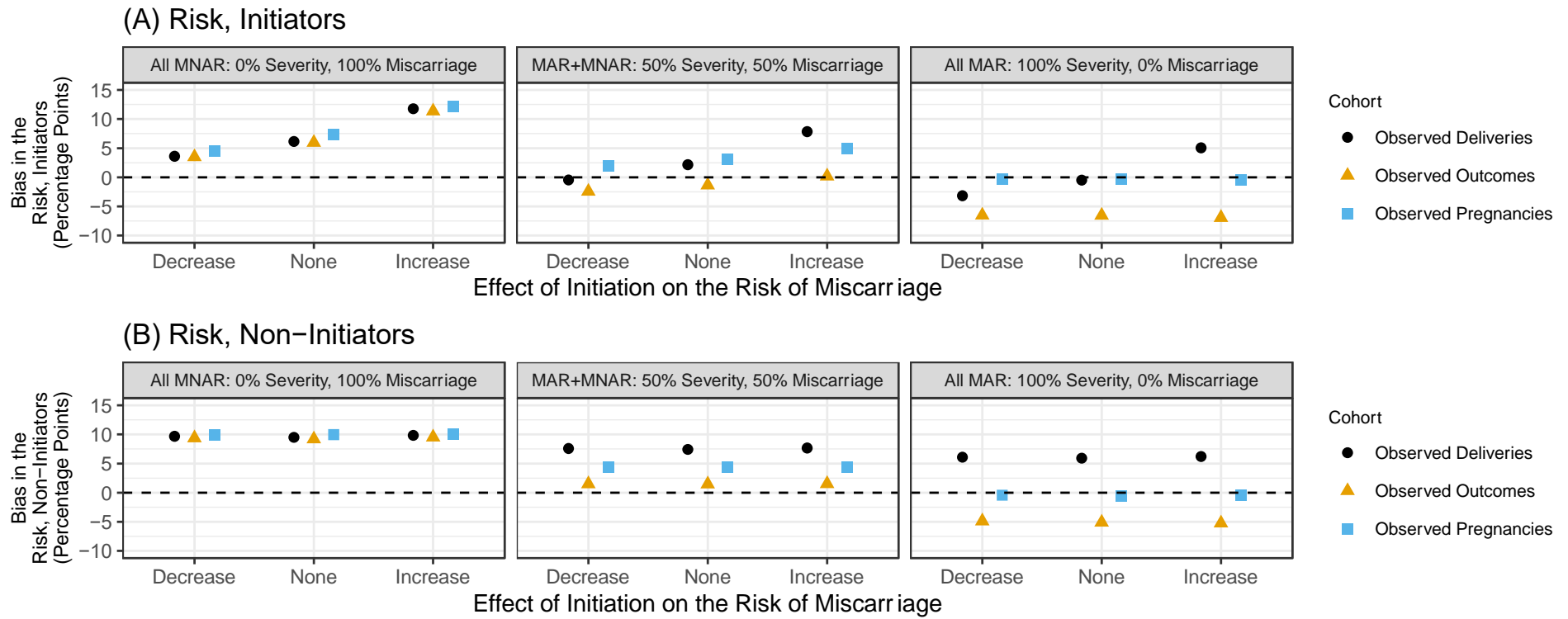

**Figure S6.** Primary and sensitivity analysis results from analytic samples where initiation did not affect the risk of preeclampsia and ~20% of pregnancies were missing outcomes. Results include bias in the (A) risk among initiators, (B) risk among non-initiators, (C) risk differences, and (D) risk ratios when outcomes were Missing Not At Random (MNAR), a mixture of MNAR and Missing at Random (MAR), or MAR. The x-axis delineates scenarios according to the effect of initiation on miscarriage (decreased, no effect on, or increased risk). The horizontal, dashed line represents no residual bias. Color distinguishes the analytic sample, and shape distinguishes whether the primary (circle) or sensitivity (triangle) analyses.

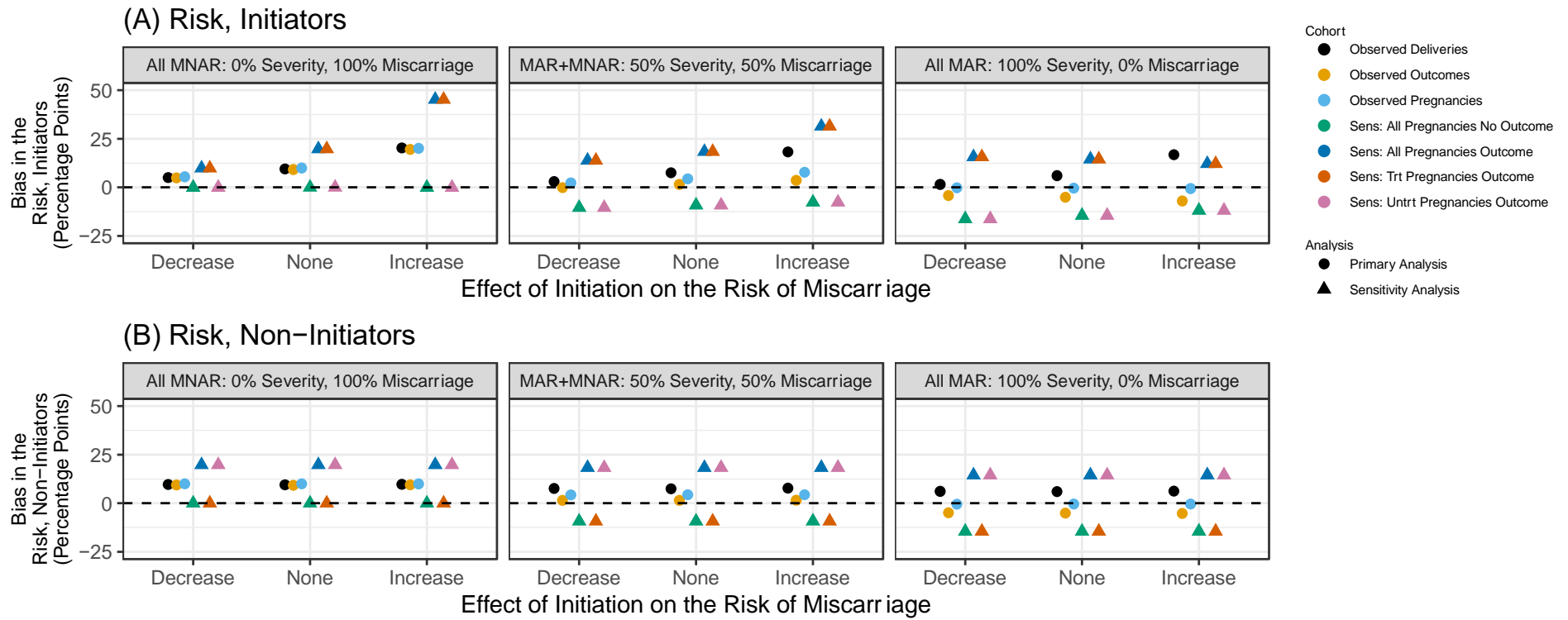

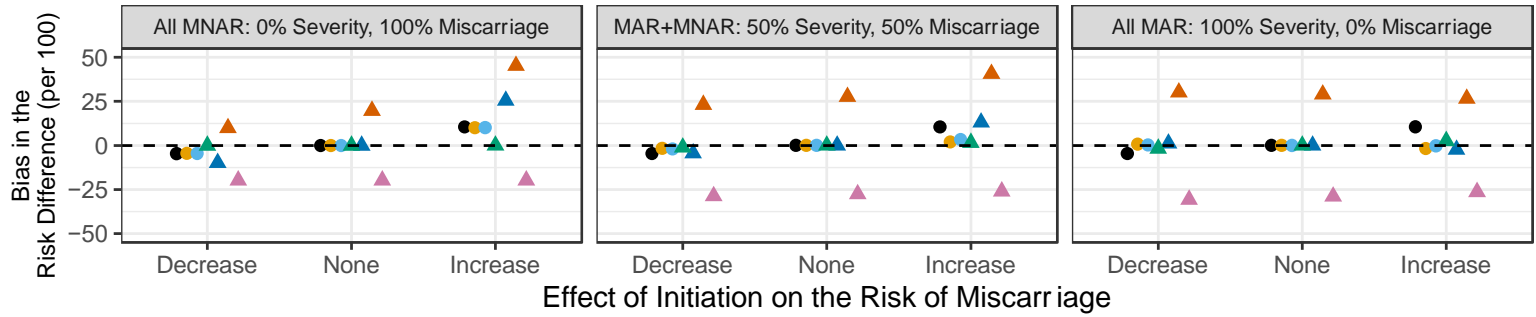

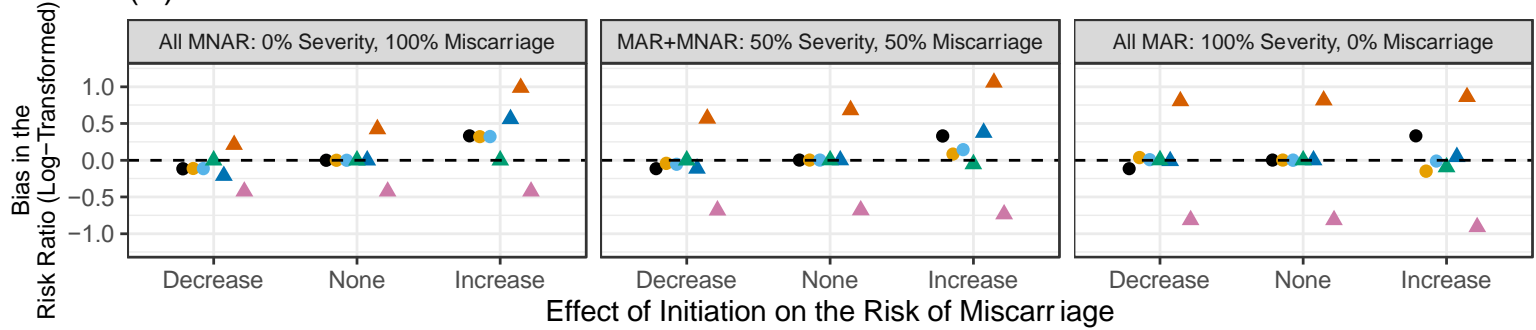

**Figure S7.** Primary and sensitivity analysis results from analytic samples where initiation decreased the risk of preeclampsia and ~20% of pregnancies were missing outcomes. Results include bias in the (A) risk among initiators, (B) risk among non-initiators, (C) risk differences, and (D) risk ratios when outcomes were Missing Not At Random (MNAR), a mixture of MNAR and Missing at Random (MAR), or MAR. The x-axis delineates scenarios according to the effect of initiation on miscarriage (decreased, no effect on, or increased risk). The horizontal, dashed line represents no residual bias. Color distinguishes the analytic sample, and shape distinguishes whether the primary (circle) or sensitivity (triangle) analyses.

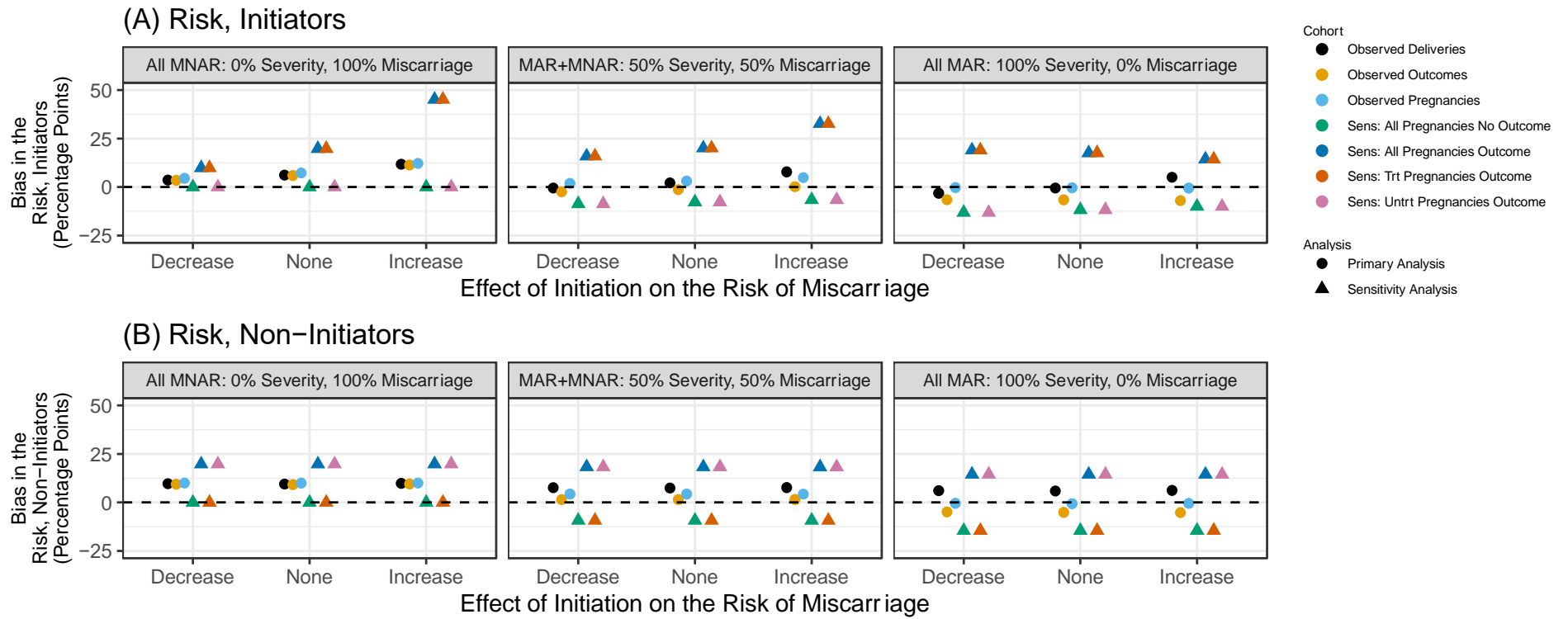

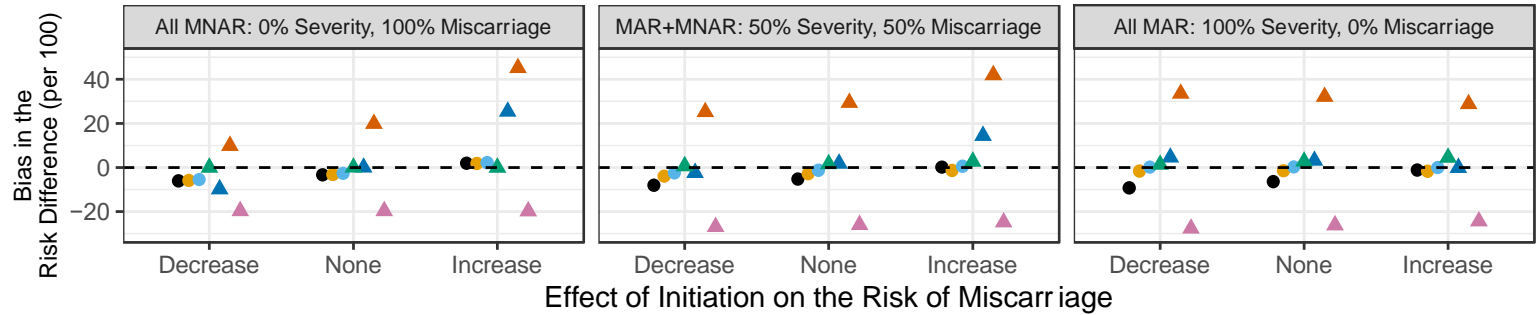

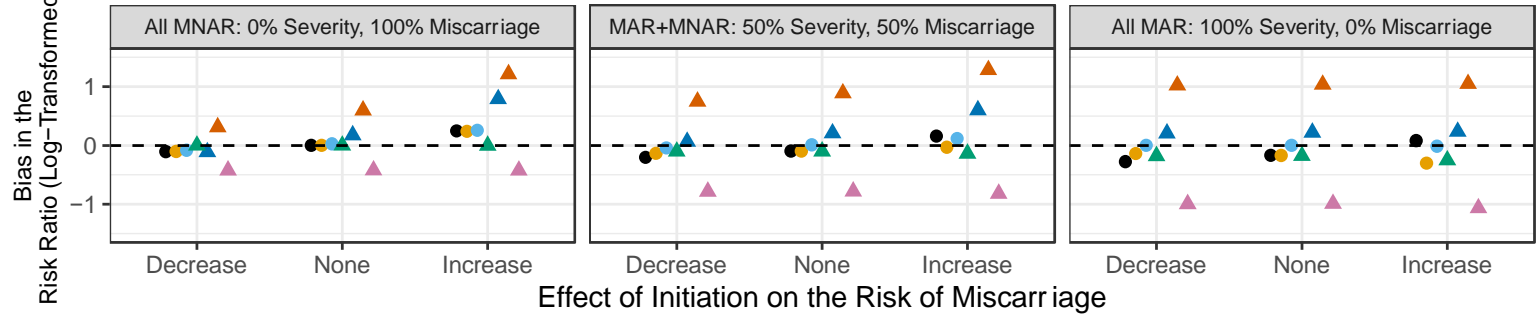

**Figure S8.** Results from scenarios where antihypertensive initiation had no direct effect on the risk of preeclampsia and ~5% of pregnancies were missing outcomes. Results include bias in the (A) risk among initiators, (B) risk among non-initiators, (C) risk difference, and (D) risk ratio when outcomes were Missing Not At Random (MNAR), a mixture of MNAR and Missing at Random (MAR), or MAR. The x-axis delineates scenarios according to the effect of initiation on miscarriage (decreased, no effect on, or increased risk). The horizontal, dashed line represents no residual bias. Analytic samples are differentiated by shape and color.

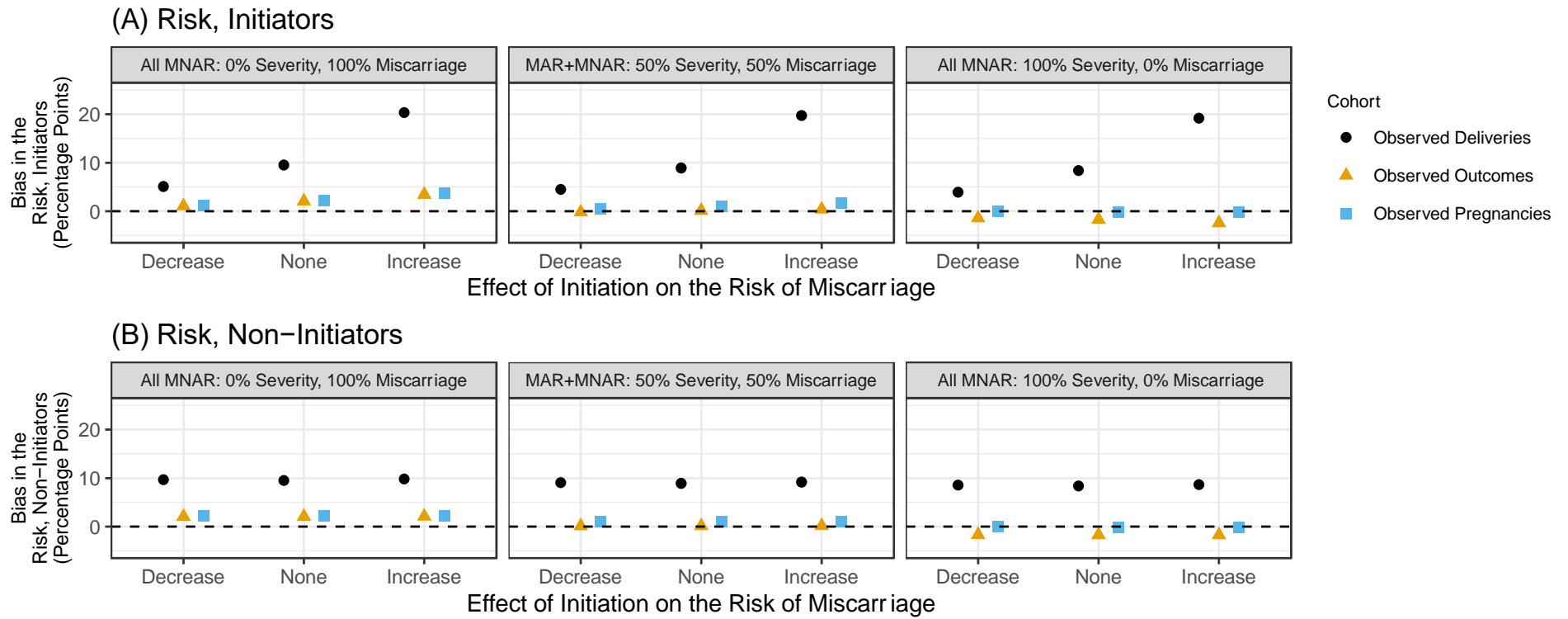

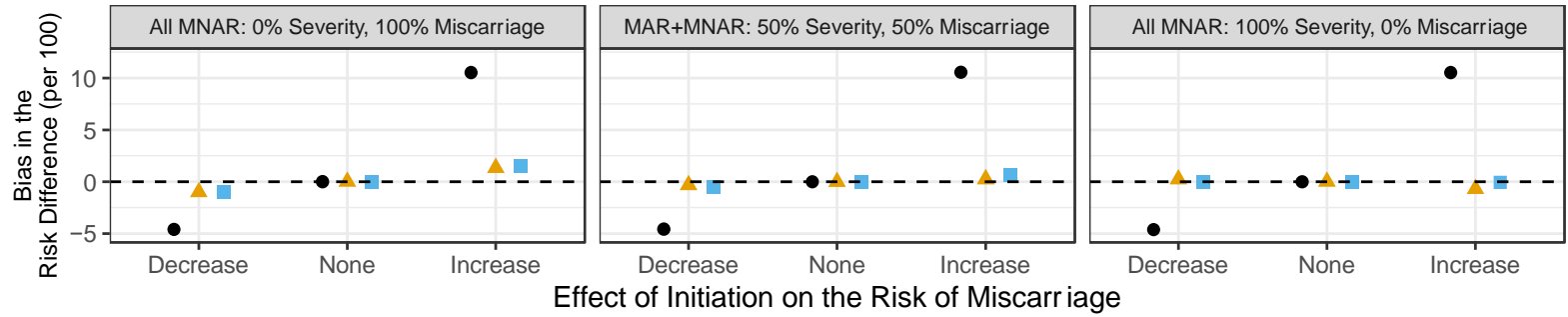

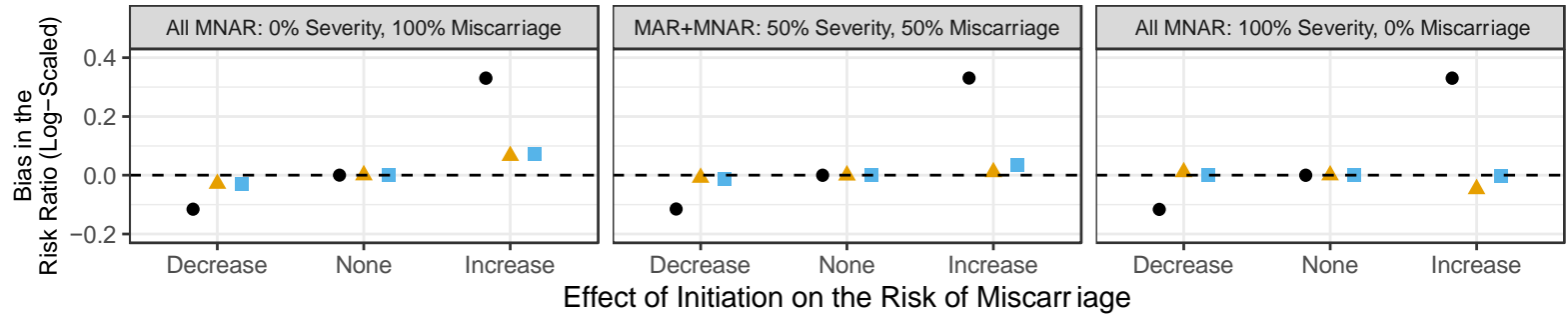

**Figure S9.** Results from scenarios where initiation decreased the risk of preeclampsia and ~5% of pregnancies were missing outcomes. Results include bias in the (A) risk among initiators, (B) risk among non-initiators, (C) risk difference, and (D) risk ratio when outcomes were Missing Not At Random (MNAR), a mixture of MNAR and Missing at Random (MAR), or MAR. The x-axis delineates scenarios according to the effect of initiation on miscarriage (decreased, no effect on, or increased risk). The horizontal, dashed line represents no residual bias. Analytic samples are differentiated by shape and color.

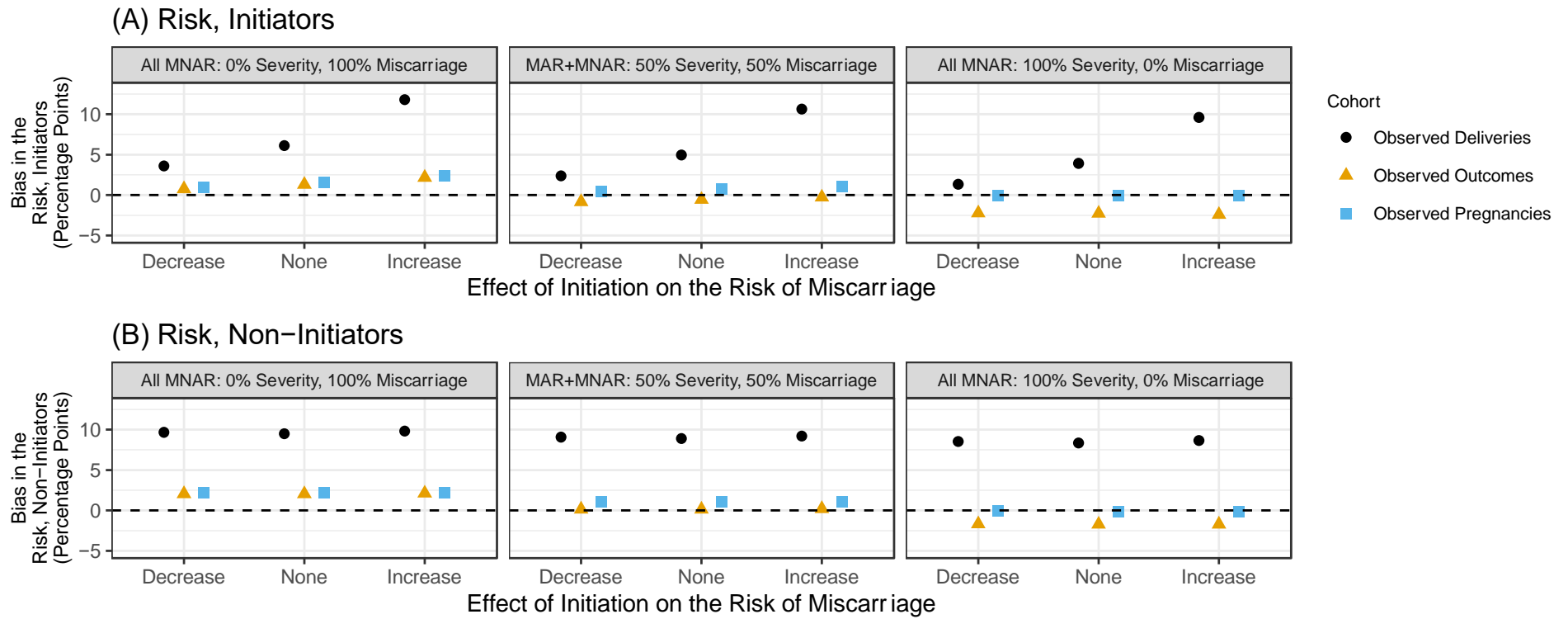

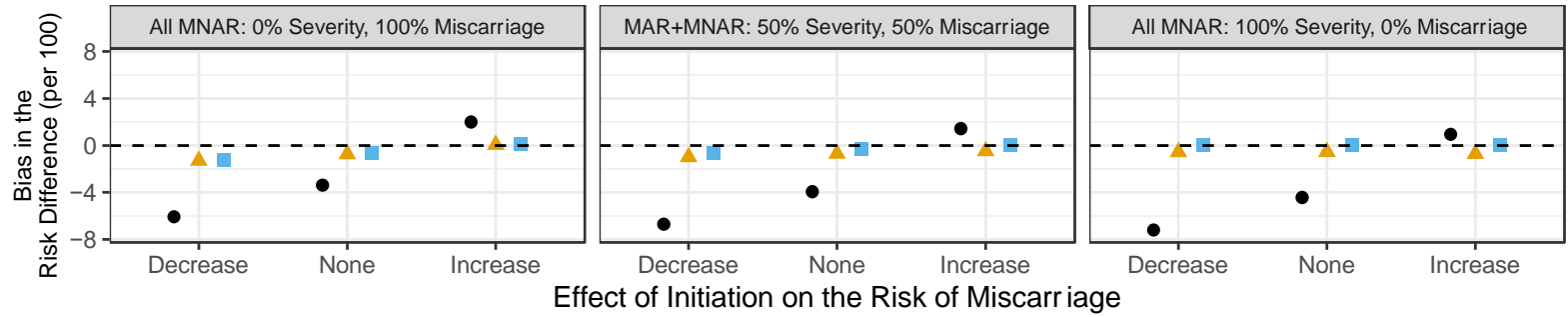

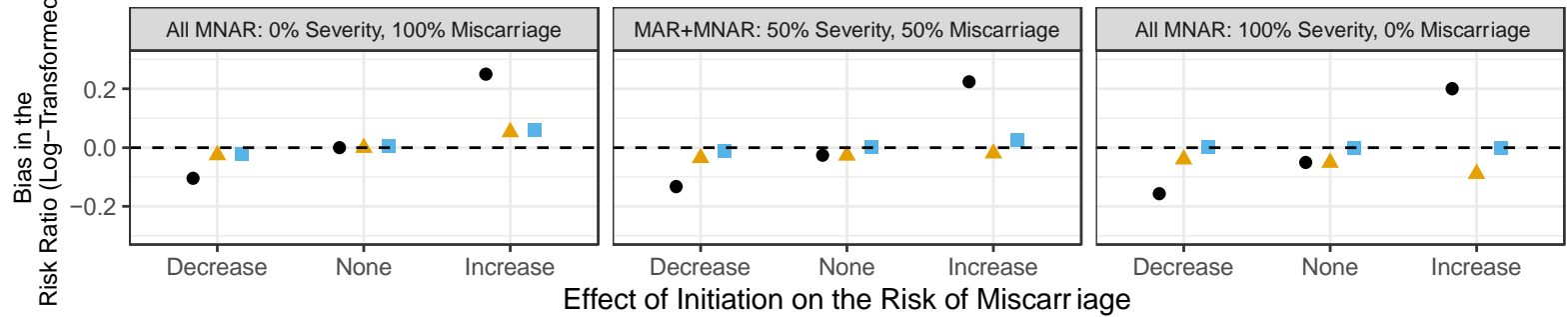

**Figure S10.** Primary and sensitivity analysis results from analytic samples where antihypertensive initiation had no direct effect on the risk of preeclampsia and ~5% of pregnancies were missing outcomes. Results include bias in the (A) risk among initiators, (B) risk among non-initiators, (C) risk differences, and (D) risk ratios when outcomes were Missing Not At Random (MNAR), a mixture of MNAR and Missing at Random (MAR), or MAR. The x-axis delineates scenarios according to the effect of initiation on miscarriage (decreased, no effect on, or increased risk). The horizontal, dashed line represents no residual bias. Color distinguishes the analytic sample, and shape distinguishes whether the primary (circle) or sensitivity (triangle) analyses.

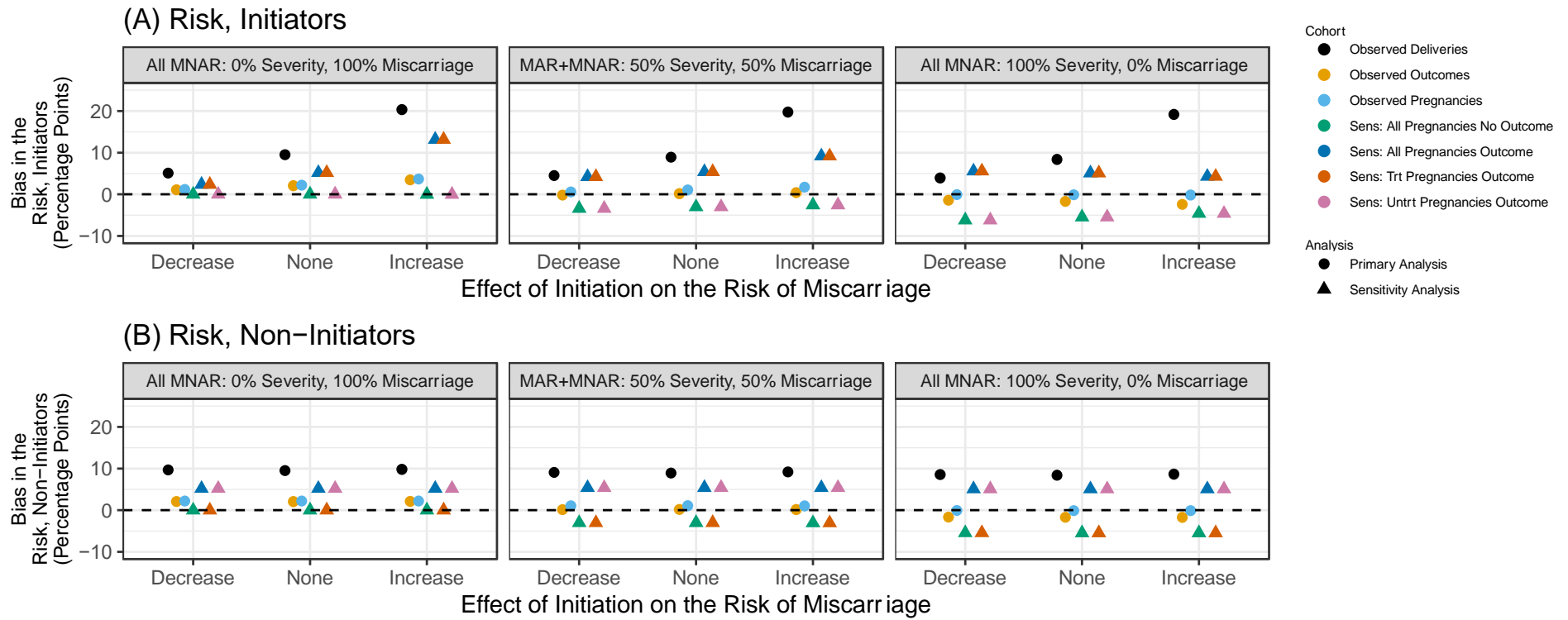

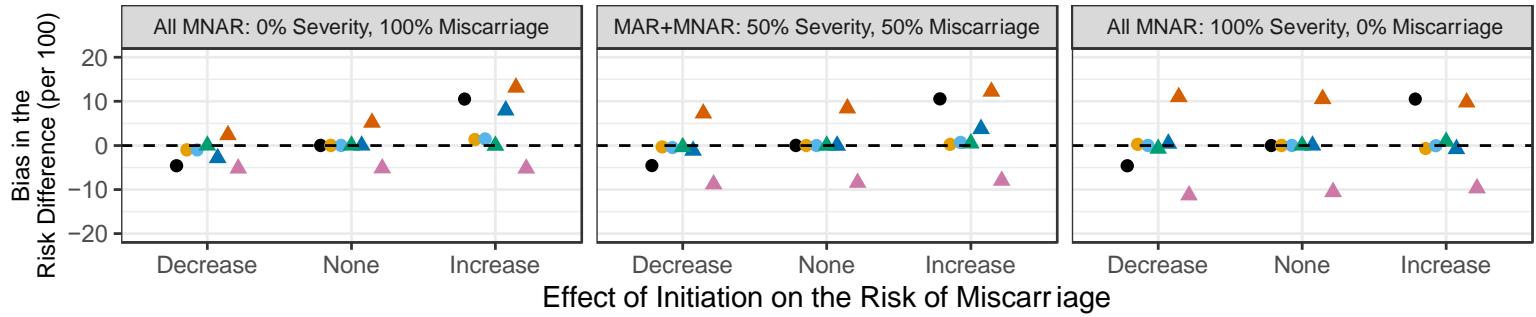
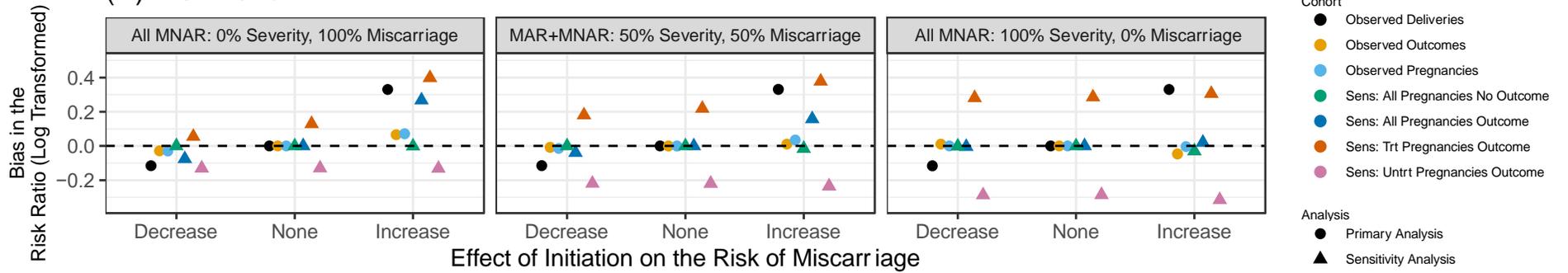

**Figure S11.** Primary and sensitivity analysis results from analytic samples where antihypertensive initiation decreased the risk of preeclampsia and ~5% of pregnancies were missing outcomes. Results include bias in the (A) risk among initiators, (B) risk among non-initiators, (C) risk differences, and (D) risk ratios when outcomes were Missing Not At Random (MNAR), a mixture of MNAR and Missing at Random (MAR), or MAR. The x-axis delineates scenarios according to the effect of initiation on miscarriage (decreased, no effect on, or increased risk). The horizontal, dashed line represents no residual bias. Color distinguishes the analytic sample, and shape distinguishes whether the primary (circle) or sensitivity (triangle) analyses.

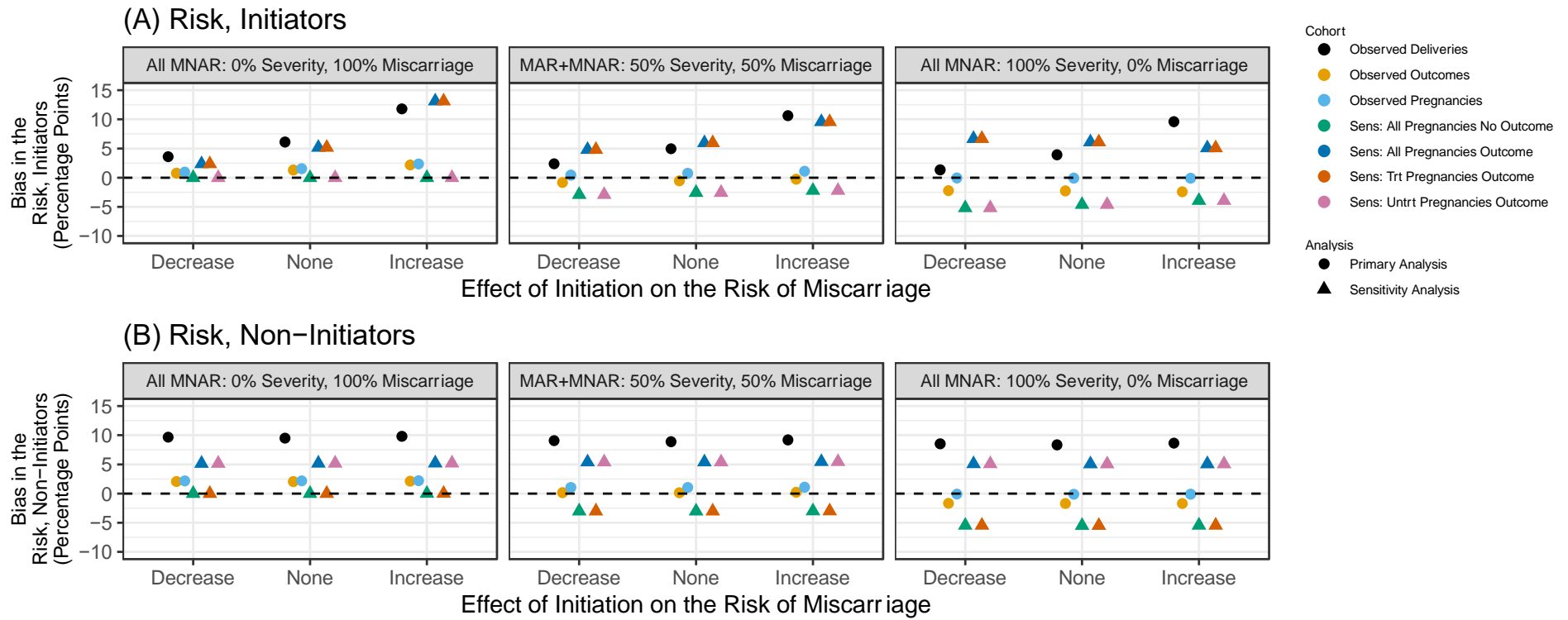

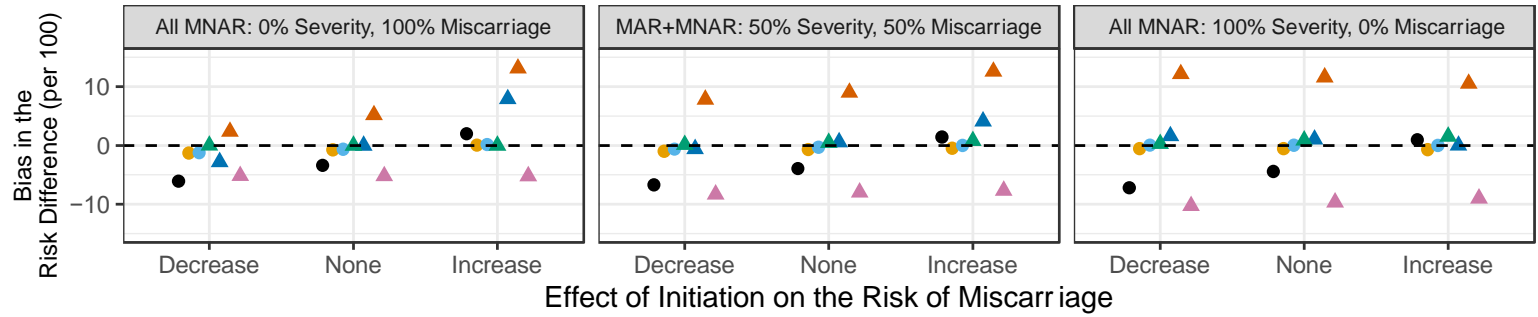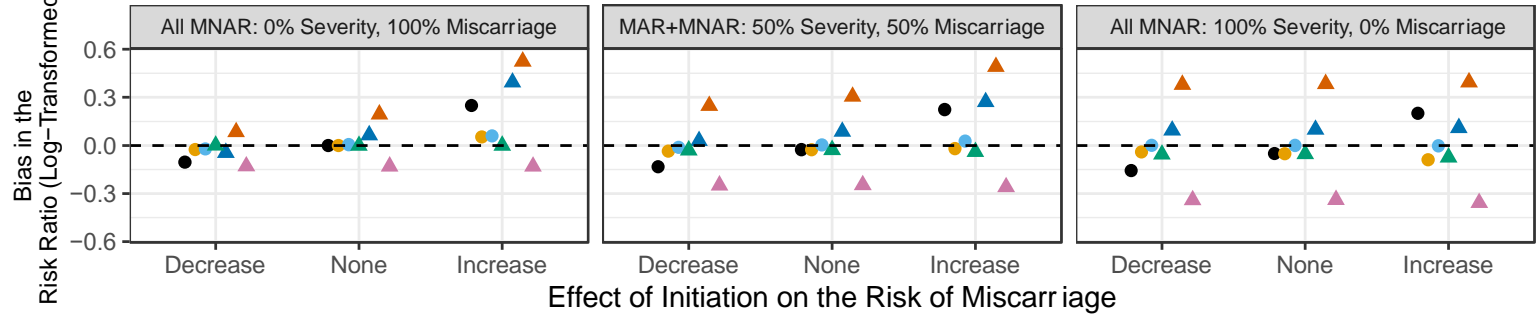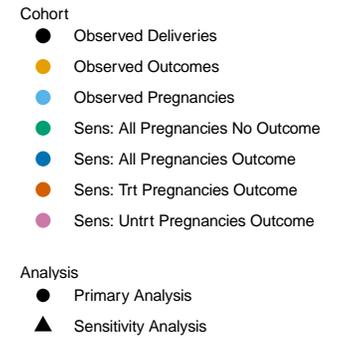

**Supplemental Tables**

**Table S1.** Tabular description of the direction of the treatment effects used to generate the 6 data target populations.

| Scenario | Direction for Effect of Initiation on Miscarriage | Direction for Effect of Initiation on Prenatal Preeclampsia |
|---|---|---|
| 1 | ⇓ | |
| 2 | — | — |
| 3 | ⇑ | |
| 4 | ⇓ | |
| 5 | — | ⇓ |
| 6 | ⇑ | |

**Table S2.** Tabular description of the 6 study samples created from each target population. Samples were defined based upon the target percentage of missingness among pregnancies that did not initiate antihypertensives.

| Study Sample | Percentage Missing an Outcome Due to Severity/Rurality | Percentage Missing an Outcome Due to Miscarriage[a] | Total Percentage of Non-Initiator Pregnancies Missing Outcomes |
|---|---|---|---|
| 1 | 0.00% | 5.00% | 5.00% |
| 2 | 2.50% | 2.50% | 5.00% |
| 3 | 5.00% | 0.00% | 5.00% |
| 4 | 0.00% | 20.00% | 20.00% |
| 5 | 10.00% | 10.00% | 20.00% |
| 6 | 20.00% | 0.00% | 20.00% |

[a] Miscarriage was defined as a fetal death that occurred prior to 20 weeks (from the pregnancy's last menstrual period).

**Table S3.** Tabular description of the marginal probabilities specified in the simulation to achieve the desired missingness distribution.

|  | Missing an Outcome Due to Severity/Rurality | | Missing an Outcome Due to Abortion | |
| --- | --- | --- | --- | --- |
| Missing Data Scenario | Desired Percentage | Marginal Probability over 32 Weeks for Balancing Intercept of Logistic Regression | Desired Percentage | Marginal Probability over 32 Weeks for Balancing Intercept of Logistic Regression[a] |
| 1 | 0.00% | 0.00% | 0.00% | 0.00% |
| 2 | 2.50% | 2.50% | 2.50% | 12.25% |
| 3 | 5.00 % | 5.00 % | 5.00% | 24.50% |
| 4 | 0.00% | 0.00% | 0.00% | 0.00% |
| 5 | 10.00% | 10.00% | 10.00% | 49.00% |
| 6 | 20.00% | 20.00% | 20.00% | 98.00% |

[a] Values were calculated by multiplying the desired percentages by 4.9.

**Table S4.** Absolute risk, risk difference per 100 pregnancies, and risk ratio estimates from each observed deliveries analytic sample.

| Goal Probability of Missingness Due to... | | Truth | | | | Observed Deliveries Analytic Sample | | | |
|---|---|---|---|---|---|---|---|---|---|
| Severity or Rurality | Miscarriage | Risk, Initiators | Risk, Non-Initiators | Risk Difference[a] | Risk Ratio | Risk, Initiators | Risk, Non-Initiators | Risk Difference[a] | Risk Ratio |
| Initiation decreases the risk of miscarriage and preeclampsia. | | | | | | | | | |
| 0.0% | 5.0% | | | | | 30.6% | 47.1% | -16.5 | 0.65 |
| 2.5% | 2.5% | | | | | 29.4% | 46.5% | -17.2 | 0.63 |
| 5.0% | 0.0% | 27.0% | 37.4% | -10.5 | 0.72 | 28.3% | 46.0% | -17.6 | 0.62 |
| 0.0% | 20.0% | | | | | 30.6% | 47.1% | -16.5 | 0.65 |
| 10.0% | 10.0% | | | | | 26.5% | 45.0% | -18.5 | 0.59 |
| 20.0% | 0.0% | | | | | 23.8% | 43.5% | -19.7 | 0.55 |
| Initiation does not affect the risk of miscarriage and decreases the risk of preeclampsia. | | | | | | | | | |
| 0.0% | 5.0% | | | | | 30.3% | 47.0% | -16.7 | 0.64 |
| 2.5% | 2.5% | | | | | 29.1% | 46.4% | -17.3 | 0.63 |
| 5.0% | 0.0% | 24.1% | 37.5% | -13.3 | 0.64 | 28.1% | 45.8% | -17.8 | 0.61 |
| 0.0% | 20.0% | | | | | 30.3% | 47.0% | -16.7 | 0.65 |
| 10.0% | 10.0% | | | | | 26.3% | 44.9% | -18.6 | 0.59 |
| 20.0% | 0.0% | | | | | 23.7% | 43.4% | -19.7 | 0.55 |
| Initiation increases the risk of miscarriage and decreases the risk of preeclampsia. | | | | | | | | | |
| 0.0% | 5.0% | | | | | 30.8% | 47.2% | -16.4 | 0.65 |
| 2.5% | 2.5% | | | | | 29.7% | 46.6% | -17.0 | 0.64 |
| 5.0% | 0.0% | 19.0% | 37.4% | -18.4 | 0.51 | 28.6% | 46.1% | -17.4 | 0.62 |
| 0.0% | 20.0% | | | | | 30.8% | 47.3% | -16.4 | 0.65 |
| 10.0% | 10.0% | | | | | 26.9% | 45.1% | -18.2 | 0.60 |
| 20.0% | 0.0% | | | | | 24.1% | 43.6% | -19.5 | 0.55 |
| Initiation decreases the risk of abortion and does not affect the risk of preeclampsia. | | | | | | | | | |
| 0.0% | 5.0% | | | | | 47.1% | 47.2% | 0.0 | 1.00 |
| 2.5% | 2.5% | | | | | 46.6% | 46.5% | 0.0 | 1.00 |
| 5.0% | 0.0% | 42.1% | 37.5% | 4.6 | 1.12 | 46.0% | 46.0% | 0.0 | 1.00 |
| 0.0% | 20.0% | | | | | 47.1% | 47.2% | -0.1 | 1.00 |
| 10.0% | 10.0% | | | | | 45.0% | 45.1% | 0.0 | 1.00 |
| 20.0% | 0.0% | | | | | 43.6% | 43.6% | 0.0 | 1.00 |
| Initiation does not affect the risk of abortion nor preeclampsia. | | | | | | | | | |
| 0.0% | 5.0% | 37.4% | 37.4% | 0.0 | 1.00 | 47.0% | 46.9% | 0.0 | 1.00 |
| 2.5% | 2.5% | | | | | 46.3% | 46.4% | 0.0 | 1.00 |
| 5.0% | 0.0% | | | | | 45.8% | 45.8% | 0.0 | 1.00 |
| 0.0% | 20.0% | | | | | 46.9% | 47.0% | -0.1 | 1.00 |
| 10.0% | 10.0% | | | | | 44.9% | 44.9% | 0.0 | 1.00 |
| 20.0% | 0.0% | | | | | 43.4% | 43.4% | 0.0 | 1.00 |
| **Initiation increases the risk of abortion and does not affect the risk of preeclampsia.** | | | | | | | | | |
| 0.0% | 5.0% | | | | | 47.2% | 47.2% | 0.0 | 1.00 |
| 2.5% | 2.5% | | | | | 46.6% | 46.6% | 0.0 | 1.00 |
| 5.0% | 0.0% | 26.9% | 37.4% | -10.5 | 0.72 | 46.1% | 46.1% | 0.0 | 1.00 |
| 0.0% | 20.0% | | | | | 47.2% | 47.2% | 0.0 | 1.00 |
| 10.0% | 10.0% | | | | | 45.2% | 45.2% | 0.0 | 1.00 |
| 20.0% | 0.0% | | | | | 43.7% | 43.7% | 0.0 | 1.00 |

[a] The risk difference represents the difference in risk per 100 pregnancies.

**Table S5.** Absolute risk, risk difference per 100 pregnancies, and risk ratio estimates from each observed outcomes analytic sample.

| Goal Probability of Missingness Due to… | | Truth | | | | Observed Outcomes Analytic Sample | | | |
|---|---|---|---|---|---|---|---|---|---|
| Severity or Rurality | Miscarriage | Risk, Initiators | Risk, Non-Initiators | Risk Difference | Risk Ratio | Risk, Initiators | Risk, Non-Initiators | Risk Difference[a] | Risk Ratio |
| Initiation decreases the risk of miscarriage and preeclampsia. | | | | | | | | | |
| 0.0% | 5.0% | | | | | 27.8% | 39.5% | -11.7 | 0.70 |
| 2.5% | 2.5% | | | | | 26.2% | 37.6% | -11.4 | 0.70 |
| 5.0% | 0.0% | 27.0% | 37.4% | -10.5 | 0.72 | 24.8% | 35.8% | -11.0 | 0.69 |
| 0.0% | 20.0% | | | | | 30.5% | 46.8% | -16.3 | 0.65 |
| 10.0% | 10.0% | | | | | 24.6% | 38.9% | -14.4 | 0.63 |
| 20.0% | 0.0% | | | | | 20.5% | 32.6% | -12.1 | 0.63 |
| Initiation does not affect the risk of miscarriage and decreases the risk of preeclampsia. | | | | | | | | | |
| 0.0% | 5.0% | | | | | 25.5% | 39.5% | -14.1 | 0.64 |
| 2.5% | 2.5% | | | | | 23.6% | 37.6% | -14.0 | 0.63 |
| 5.0% | 0.0% | 24.1% | 37.5% | -13.3 | 0.64 | 21.9% | 35.8% | -13.9 | 0.61 |
| 0.0% | 20.0% | | | | | 30.1% | 46.7% | -16.6 | 0.65 |
| 10.0% | 10.0% | | | | | 22.8% | 38.9% | -16.1 | 0.59 |
| 20.0% | 0.0% | | | | | 17.6% | 32.4% | -14.8 | 0.54 |
| Initiation increases the risk of miscarriage and decreases the risk of preeclampsia. | | | | | | | | | |
| 0.0% | 5.0% | | | | | 21.2% | 39.6% | -18.3 | 0.54 |
| 2.5% | 2.5% | | | | | 18.8% | 37.6% | -18.9 | 0.50 |
| 5.0% | 0.0% | 19.0% | 37.4% | -18.4 | 0.51 | 16.6% | 35.7% | -19.1 | 0.47 |
| 0.0% | 20.0% | | | | | 30.4% | 47.0% | -16.6 | 0.65 |
| 10.0% | 10.0% | | | | | 19.2% | 39.0% | -19.7 | 0.49 |
| 20.0% | 0.0% | | | | | 12.1% | 32.2% | -20.1 | 0.38 |
| Initiation decreases the risk of miscarriage and does not affect the risk of preeclampsia. | | | | | | | | | |
| 0.0% | 5.0% | | | | | 43.1% | 39.5% | 3.6 | 1.09 |
| 2.5% | 2.5% | | | | | 41.9% | 37.6% | 4.3 | 1.11 |
| 5.0% | 0.0% | 42.1% | 37.5% | 4.6 | 1.12 | 40.6% | 35.8% | 4.8 | 1.13 |
| 0.0% | 20.0% | | | | | 46.9% | 46.9% | 0.1 | 1.00 |
| 10.0% | 10.0% | | | | | 41.9% | 39.0% | 2.9 | 1.07 |
| 20.0% | 0.0% | | | | | 37.9% | 32.5% | 5.3 | 1.16 |
| Initiation does not affect the risk of miscarriage nor preeclampsia. | | | | | | | | | |
| 0.0% | 5.0% | 37.4% | 37.4% | 0.0 | 1.00 | 39.5% | 39.5% | 0.0 | 1.00 |
| 2.5% | 2.5% | | | | | 37.6% | 37.6% | 0.0 | 1.00 |
| 5.0% | 0.0% | | | | | 35.7% | 35.7% | 0.0 | 1.00 |
| 0.0% | 20.0% | | | | | 46.6% | 46.7% | -0.1 | 1.00 |
| 10.0% | 10.0% | | | | | 38.9% | 38.9% | 0.0 | 1.00 |
| 20.0% | 0.0% | | | | | 32.4% | 32.4% | 0.0 | 1.00 |
| **Initiation increases the risk of miscarriage and does not affect the risk of preeclampsia.** | | | | | | | | | |
| 0.0% | 5.0% | | | | | 30.4% | 39.5% | -9.2 | 0.77 |
| 2.5% | 2.5% | | | | | 27.3% | 37.6% | -10.3 | 0.73 |
| 5.0% | 0.0% | 26.9% | 37.4% | -10.5 | 0.72 | 24.5% | 35.7% | -11.2 | 0.69 |
| 0.0% | 20.0% | | | | | 46.4% | 46.9% | -0.5 | 0.99 |
| 10.0% | 10.0% | | | | | 30.5% | 39.0% | -8.5 | 0.78 |
| 20.0% | 0.0% | | | | | 19.9% | 32.2% | -12.3 | 0.62 |

<sup>a</sup> The risk difference represents the difference in risk per 100 pregnancies.

Table S6. Absolute risk, risk difference per 100 pregnancies, and risk ratio estimates from each observed pregnancies analytic sample.

| Goal Probability of Missingness Due to… | | Truth | | | | Observed Pregnancies Analytic Sample | | | |
|---|---|---|---|---|---|---|---|---|---|
| Severity or Rurality | Miscarriage | Risk, Initiators | Risk, Non-Initiators | Risk Difference | Risk Ratio | Risk, Initiators | Risk, Non-Initiators | Risk Difference[a] | Risk Ratio |
| Initiation decreases the risk of miscarriage and preeclampsia. | | | | | | | | | |
| 0.0% | 5.0% | | | | | 27.9% | 39.6% | -11.7 | 0.71 |
| 2.5% | 2.5% | | | | | 27.4% | 38.5% | -11.1 | 0.71 |
| 5.0% | 0.0% | 27.0% | 37.4% | -10.5 | 0.72 | 26.9% | 37.3% | -10.4 | 0.72 |
| 0.0% | 20.0% | | | | | 31.5% | 47.4% | -15.9 | 0.66 |
| 10.0% | 10.0% | | | | | 28.9% | 41.8% | -12.9 | 0.69 |
| 20.0% | 0.0% | | | | | 26.7% | 37.0% | -10.3 | 0.72 |
| Initiation does not affect the risk of miscarriage and decreases the risk of preeclampsia. | | | | | | | | | |
| 0.0% | 5.0% | | | | | 25.7% | 39.7% | -14.0 | 0.65 |
| 2.5% | 2.5% | | | | | 24.9% | 38.5% | -13.6 | 0.65 |
| 5.0% | 0.0% | 24.1% | 37.5% | -13.3 | 0.64 | 24.1% | 37.3% | -13.3 | 0.64 |
| 0.0% | 20.0% | | | | | 31.4% | 47.4% | -16.0 | 0.66 |
| 10.0% | 10.0% | | | | | 27.2% | 41.8% | -14.6 | 0.65 |
| 20.0% | 0.0% | | | | | 23.8% | 36.8% | -13.1 | 0.65 |
| Initiation increases the risk of miscarriage and decreases the risk of preeclampsia. | | | | | | | | | |
| 0.0% | 5.0% | | | | | 21.4% | 39.6% | -18.2 | 0.54 |
| 2.5% | 2.5% | | | | | 20.1% | 38.5% | -18.4 | 0.52 |
| 5.0% | 0.0% | 19.0% | 37.4% | -18.4 | 0.51 | 18.9% | 37.3% | -18.4 | 0.51 |
| 0.0% | 20.0% | | | | | 31.2% | 47.4% | -16.2 | 0.66 |
| 10.0% | 10.0% | | | | | 23.9% | 41.7% | -17.8 | 0.57 |
| 20.0% | 0.0% | | | | | 18.5% | 36.9% | -18.4 | 0.50 |
| Initiation decreases the risk of abortion and does not affect the risk of preeclampsia. | | | | | | | | | |
| 0.0% | 5.0% | | | | | 43.2% | 39.7% | 3.6 | 1.09 |
| 2.5% | 2.5% | | | | | 42.6% | 38.5% | 4.1 | 1.11 |
| 5.0% | 0.0% | 42.1% | 37.5% | 4.6 | 1.12 | 42.0% | 37.4% | 4.6 | 1.12 |
| 0.0% | 20.0% | | | | | 47.5% | 47.4% | 0.1 | 1.00 |
| 10.0% | 10.0% | | | | | 44.4% | 41.8% | 2.6 | 1.06 |
| 20.0% | 0.0% | | | | | 41.8% | 37.0% | 4.9 | 1.13 |
| Initiation does not affect the risk of abortion nor preeclampsia. | | | | | | | | | |
| 0.0% | 5.0% | 37.4% | 37.4% | 0.0 | 1.00 | 39.6% | 39.6% | 0.0 | 1.00 |
| 2.5% | 2.5% | | | | | 38.5% | 38.5% | 0.0 | 1.00 |
| 5.0% | 0.0% | | | | | 37.3% | 37.3% | 0.0 | 1.00 |
| 0.0% | 20.0% | | | | | 47.4% | 47.4% | -0.1 | 1.00 |
| 10.0% | 10.0% | | | | | 41.8% | 41.8% | 0.0 | 1.00 |
| 20.0% | 0.0% | | | | | 37.0% | 37.1% | 0.0 | 1.00 |
| **Initiation increases the risk of abortion and does not affect the risk of preeclampsia.** | | | | | | | | | |
| 0.0% | 5.0% | | | | | 30.6% | 39.6% | -9.0 | 0.77 |
| 2.5% | 2.5% | | | | | 28.6% | 38.4% | -9.8 | 0.74 |
| 5.0% | 0.0% | 26.9% | 37.4% | -10.5 | 0.72 | 26.7% | 37.3% | -10.6 | 0.72 |
| 0.0% | 20.0% | | | | | 47.0% | 47.4% | -0.4 | 0.99 |
| 10.0% | 10.0% | | | | | 34.7% | 41.8% | -7.1 | 0.83 |
| 20.0% | 0.0% | | | | | 26.3% | 37.0% | -10.7 | 0.71 |

[a] The risk difference represents the difference in risk per 100 pregnancies.

**Table S7.** Descriptive statistics of the simulated pregnancies included in each study sample where ~20% of pregnancies were missing outcomes.

| Goal Percentage Missing Due to… | | Treatment | Number (%) Missing Due to… | | Number of Pregnancies Included in the Study Sample by Severity N (%) | | | True Pregnancy Outcomes N (%) | | |
|---|---|---|---|---|---|---|---|---|---|---|
| Severity or Rurality | Miscarriage | | Severity or Rurality | Miscarriage | Low | Moderate | High | Miscarriage[a] | Stillbirth[b] | Live birth |
| **Initiation decreases the risk of miscarriage and preeclampsia.** | | | | | | | | | | |
| 0.0% | 20.0% | Initiate | 0 (0%) | 652,942 (13%) | 833,873 (17%) | 1,666,480 (33%) | 2,500,426 (50%) | 672,597 (13%) | 63,424 (1%) | 4,262,985 (85%) |
| | | Not initiate | 0 (0%) | 866,190 (17%) | 2,500,605 (50%) | 1,666,900 (33%) | 831,716 (17%) | 888,471 (18%) | 67,613 (1%) | 4,044,910 (81%) |
| 10.0% | 10.0% | Initiate | 1,337,443 (27%) | 280,951 (6%) | 834,754 (17%) | 1,667,249 (33%) | 2,498,806 (50%) | 671,952 (13%) | 63,410 (1%) | 4,263,979 (85%) |
| | | Not initiate | 611,365 (12%) | 425,692 (9%) | 2,499,724 (50%) | 1,666,131 (33%) | 833,336 (17%) | 888,921 (18%) | 67,788 (1%) | 4,043,950 (81%) |
| 20.0% | 0.0% | Initiate | 2,063,804 (41%) | 0 (0%) | 832,934 (17%) | 1,665,873 (33%) | 2,498,748 (50%) | 672,292 (13%) | 63,772 (1%) | 4,263,604 (85%) |
| | | Not initiate | 1,030,878 (21%) | 0 (0%) | 2,501,544 (50%) | 1,667,507 (33%) | 833,394 (17%) | 889,125 (18%) | 67,594 (1%) | 4,043,613 (81%) |
| **Initiation does not affect the risk of miscarriage and decreases the risk of preeclampsia.** | | | | | | | | | | |
| 0.0% | 20.0% | Initiate | 0 (0%) | 1,108,211 (22%) | 832,202 (17%) | 1,666,535 (33%) | 2,498,467 (50%) | 1,136,279 (23%) | 56,963 (1%) | 3,803,962 (76%) |
| | | Not initiate | 0 (0%) | 867,756 (17%) | 2,501,943 (50%) | 1,668,773 (33%) | 832,080 (17%) | 889,123 (18%) | 67,616 (1%) | 4,046,057 (81%) |
| 10.0% | 10.0% | Initiate | 1,211,873 (24%) | 521,433 (10%) | 833,764 (17%) | 1,666,538 (33%) | 2,498,961 (50%) | 1,135,239 (23%) | 56,858 (1%) | 3,807,166 (76%) |
| | | Not initiate | 611,345 (12%) | 426,437 (9%) | 2,500,381 (50%) | 1,668,770 (33%) | 831,586 (17%) | 888,643 (18%) | 67,399 (1%) | 4,044,695 (81%) |
| 20.0% | 0.0% | Initiate | 1,886,169 (38%) | 0 (0%) | 834,512 (17%) | 1,667,537 (33%) | 2,498,290 (50%) | 1,134,553 (23%) | 56,989 (1%) | 3,808,797 (76%) |
| | | Not initiate | 1,028,594 (21%) | 0 (0%) | 2,499,633 (50%) | 1,667,771 (33%) | 832,257 (17%) | 888,055 (18%) | 67,821 (1%) | 4,043,785 (81%) |
| **Initiation increases the risk of miscarriage and decreases the risk of preeclampsia.** | | | | | | | | | | |
| 0.0% | 20.0% | Initiate | 0 (0%) | 2,038,896 (41%) | 833,011 (17%) | 1,667,726 (33%) | 2,499,616 (50%) | 2,081,254 (42%) | 44,648 (1%) | 2,874,451 (57%) |
| | | Not initiate | 0 (0%) | 868,457 (17%) | 2,498,723 (50%) | 1,667,698 (33%) | 833,226 (17%) | 889,801 (18%) | 67,776 (1%) | 4,042,070 (81%) |
| 10.0% | 10.0% | Initiate | 1,048,268 (21%) | 1,047,191 (21%) | 833,152 (17%) | 1,668,069 (33%) | 2,500,230 (50%) | 2,082,341 (42%) | 44,865 (1%) | 2,874,245 (57%) |
| | | Not initiate | 609,849 (12%) | 425,779 (9%) | 2,498,582 (50%) | 1,667,355 (33%) | 832,612 (17%) | 888,624 (18%) | 67,531 (1%) | 4,042,394 (81%) |
| 20.0% | 0.0% | Initiate | 1,627,219 (33%) | 0 (0%) | 832,562 (17%) | 1,668,460 (33%) | 2,499,586 (50%) | 2,081,842 (42%) | 44,644 (1%) | 2,874,122 (57%) |
| | | Not initiate | 1,027,859 (21%) | 0 (0%) | 2,499,172 (50%) | 1,666,964 (33%) | 833,256 (17%) | 889,439 (18%) | 67,644 (1%) | 4,042,309 (81%) |

**Initiation decreases the risk of abortion and does not affect the risk of preeclampsia.**

| | | | | | | | | | | |
|---|---|---|---|---|---|---|---|---|---|---|
| 0.0% | 20.0% | Initiate | 0 (0%) | 652,323 (13%) | 832,900 (17%) | 1,668,820 (33%) | 2,500,411 (50%) | 671,748 (13%) | 87,300 (2%) | 4,243,083 (85%) |
| | | Not initiate | 0 (0%) | 867,182 (17%) | 2,497,359 (50%) | 1,668,018 (33%) | 832,492 (17%) | 888,339 (18%) | 67,398 (1%) | 4,042,132 (81%) |
| 10.0% | 10.0% | Initiate | 1,318,550 (26%) | 280,976 (6%) | 831,811 (17%) | 1,669,128 (33%) | 2,497,932 (50%) | 672,287 (13%) | 87,486 (2%) | 4,239,098 (85%) |
| | | Not initiate | 611,664 (12%) | 426,663 (9%) | 2,498,448 (50%) | 1,667,710 (33%) | 834,971 (17%) | 889,110 (18%) | 67,377 (1%) | 4,044,642 (81%) |
| 20.0% | 0.0% | Initiate | 2,045,506 (41%) | 0 (0%) | 833,265 (17%) | 1,669,718 (33%) | 2,499,793 (50%) | 671,964 (13%) | 87,204 (2%) | 4,243,608 (85%) |
| | | Not initiate | 1,029,202 (21%) | 0 (0%) | 2,496,994 (50%) | 1,667,120 (33%) | 833,110 (17%) | 889,105 (18%) | 67,215 (1%) | 4,040,904 (81%) |

**Initiation does not affect the risk of abortion or preeclampsia.**

| | | | | | | | | | | |
|---|---|---|---|---|---|---|---|---|---|---|
| 0.0% | 20.0% | Initiate | 0 (0%) | 1,108,266 (22%) | 833,862 (17%) | 1,664,398 (33%) | 2,502,550 (50%) | 1,136,678 (23%) | 78,154 (2%) | 3,785,978 (76%) |
| | | Not initiate | 0 (0%) | 869,015 (17%) | 2,499,128 (50%) | 1,667,996 (33%) | 832,066 (17%) | 890,045 (18%) | 67,932 (1%) | 4,041,213 (81%) |
| 10.0% | 10.0% | Initiate | 1,199,228 (24%) | 522,999 (10%) | 834,709 (17%) | 1,666,254 (33%) | 2,501,363 (50%) | 1,137,971 (23%) | 78,048 (2%) | 3,786,307 (76%) |
| | | Not initiate | 610,528 (12%) | 426,463 (9%) | 2,498,281 (50%) | 1,666,140 (33%) | 833,253 (17%) | 889,760 (18%) | 67,521 (1%) | 4,040,393 (81%) |
| 20.0% | 0.0% | Initiate | 1,867,374 (37%) | 0 (0%) | 834,165 (17%) | 1,665,499 (33%) | 2,500,930 (50%) | 1,136,962 (23%) | 78,321 (2%) | 3,785,311 (76%) |
| | | Not initiate | 1,028,177 (21%) | 0 (0%) | 2,498,825 (50%) | 1,666,895 (33%) | 833,686 (17%) | 889,625 (18%) | 67,650 (1%) | 4,042,131 (81%) |

**Initiation increases the risk of abortion and does not affect the risk of preeclampsia.**

| | | | | | | | | | | |
|---|---|---|---|---|---|---|---|---|---|---|
| 0.0% | 20.0% | Initiate | 0 (0%) | 2,040,782 (41%) | 833,710 (17%) | 1,664,790 (33%) | 2,498,623 (50%) | 2,083,438 (42%) | 61,778 (1%) | 2,851,907 (57%) |
| | | Not initiate | 0 (0%) | 868,582 (17%) | 2,499,917 (50%) | 1,668,849 (33%) | 834,111 (17%) | 889,612 (18%) | 67,525 (1%) | 4,045,740 (81%) |
| 10.0% | 10.0% | Initiate | 1,034,655 (21%) | 1,050,648 (21%) | 833,102 (17%) | 1,665,927 (33%) | 2,499,657 (50%) | 2,082,551 (42%) | 61,860 (1%) | 2,854,275 (57%) |
| | | Not initiate | 611,049 (12%) | 427,521 (9%) | 2,500,525 (50%) | 1,667,712 (33%) | 833,077 (17%) | 890,970 (18%) | 67,536 (1%) | 4,042,808 (81%) |
| 20.0% | 0.0% | Initiate | 1,610,066 (32%) | 0 (0%) | 833,165 (17%) | 1,667,263 (33%) | 2,498,648 (50%) | 2,082,877 (42%) | 61,960 (1%) | 2,854,239 (57%) |
| | | Not initiate | 1,030,070 (21%) | 0 (0%) | 2,500,462 (50%) | 1,666,376 (33%) | 834,086 (17%) | 890,401 (18%) | 67,523 (1%) | 4,043,000 (81%) |

[a] Defined as fetal death prior to 20 weeks of gestation from LMP (18 from conception).
[b] Defined as fetal death at or after 20 weeks of gestation from LMP (18 from conception).

**Table S8.** Descriptive statistics of the simulated pregnancies included in each study sample where ~5% of pregnancies were missing outcomes.

| Goal Percentage Missing Due to… | | Treatment | Number (%) Missing Due to… | | Number of Pregnancies Included in the Study Sample by Severity N (%) | | | True Pregnancy Outcomes N (%) | | |
|---|---|---|---|---|---|---|---|---|---|---|
| Severity or Rurality | Miscarriage | | Severity or Rurality | Miscarriage | Low | Moderate | High | Miscarriage[a] | Stillbirth[b] | Live birth |
| **Initiation decreases the risk of miscarriage and preeclampsia.** | | | | | | | | | | |
| 0.0% | 5.0% | Initiate | 0 (0%) | 155,811 (3%) | 833,518 (17%) | 1,665,028 (33%) | 2,500,460 (50%) | 672,597 (13%) | 63,424 (1%) | 4,262,985 (85%) |
| | | Not initiate | 0 (0%) | 230,117 (5%) | 2,500,960 (50%) | 1,668,352 (33%) | 831,682 (17%) | 888,471 (18%) | 67,613 (1%) | 4,044,910 (81%) |
| 2.5% | 2.5% | Initiate | 443,572 (9%) | 77,816 (2%) | 833,500 (17%) | 1,666,820 (33%) | 2,499,021 (50%) | 671,952 (13%) | 63,410 (1%) | 4,263,979 (85%) |
| | | Not initiate | 185,586 (4%) | 117,446 (2%) | 2,500,978 (50%) | 1,666,560 (33%) | 833,121 (17%) | 888,921 (18%) | 67,788 (1%) | 4,043,950 (81%) |
| 5.0% | 0.0% | Initiate | 797,415 (16%) | 0 (0%) | 834,354 (17%) | 1,666,218 (33%) | 2,499,096 (50%) | 672,292 (13%) | 63,772 (1%) | 4,263,604 (85%) |
| | | Not initiate | 344,969 (7%) | 0 (0%) | 2,500,124 (50%) | 1,667,162 (33%) | 833,046 (17%) | 889,125 (18%) | 67,594 (1%) | 4,043,613 (81%) |
| **Initiation does not affect the risk of miscarriage and decreases the risk of preeclampsia.** | | | | | | | | | | |
| 0.0% | 5.0% | Initiate | 0 (0%) | 288,933 (6%) | 833,327 (17%) | 1,666,564 (33%) | 2,498,955 (50%) | 1,136,399 (23%) | 56,894 (1%) | 3,805,553 (76%) |
| | | Not initiate | 0 (0%) | 230,514 (5%) | 2,500,818 (50%) | 1,668,744 (33%) | 831,592 (17%) | 888,107 (18%) | 67,521 (1%) | 4,045,526 (81%) |
| 2.5% | 2.5% | Initiate | 399,686 (8%) | 146,890 (3%) | 833,613 (17%) | 1,668,725 (33%) | 2,499,165 (50%) | 1,136,635 (23%) | 57,187 (1%) | 3,807,681 (76%) |
| | | Not initiate | 185,748 (4%) | 117,821 (2%) | 2,500,532 (50%) | 1,666,583 (33%) | 831,382 (17%) | 887,541 (18%) | 67,632 (1%) | 4,043,324 (81%) |
| 5.0% | 0.0% | Initiate | 721,181 (14%) | 0 (0%) | 832,737 (17%) | 1,667,140 (33%) | 2,498,176 (50%) | 1,135,212 (23%) | 56,743 (1%) | 3,806,098 (76%) |
| | | Not initiate | 345,245 (7%) | 0 (0%) | 2,501,408 (50%) | 1,668,168 (33%) | 832,371 (17%) | 889,217 (18%) | 67,566 (1%) | 4,045,164 (81%) |
| **Initiation increases the risk of miscarriage and decreases the risk of preeclampsia.** | | | | | | | | | | |

| | | | | | | | | | | | |
|---|---|---|---|---|---|---|---|---|---|---|---|
| 0.0% | 5.0% | Initiate | 0 (0%) | 579,813 (12%) | | 832,813 (17%) | 1,666,292 (33%) | 2,500,676 (50%) | 2,080,444 (42%) | 44,930 (1%) | 2,874,407 (57%) |
| | | Not initiate | 0 (0%) | 230,768 (5%) | | 2,498,921 (50%) | 1,669,132 (33%) | 832,166 (17%) | 888,670 (18%) | 67,671 (1%) | 4,043,878 (81%) |
| 2.5% | 2.5% | Initiate | 344,716 (7%) | 299,908 (6%) | | 832,086 (17%) | 1,667,130 (33%) | 2,500,035 (50%) | 2,081,503 (42%) | 44,617 (1%) | 2,873,131 (57%) |
| | | Not initiate | 186,472 (4%) | 118,352 (2%) | | 2,499,648 (50%) | 1,668,294 (33%) | 832,807 (17%) | 887,550 (18%) | 67,462 (1%) | 4,045,737 (81%) |
| 5.0% | 0.0% | Initiate | 624,074 (12%) | 0 (0%) | | 833,874 (17%) | 1,667,210 (33%) | 2,501,375 (50%) | 2,081,219 (42%) | 44,761 (1%) | 2,876,479 (58%) |
| | | Not initiate | 343,886 (7%) | 0 (0%) | | 2,497,860 (50%) | 1,668,214 (33%) | 831,467 (17%) | 890,017 (18%) | 67,655 (1%) | 4,039,869 (81%) |

**Initiation decreases the risk of abortion and does not affect the risk of preeclampsia.**

| | | | | | | | | | | | |
|---|---|---|---|---|---|---|---|---|---|---|---|
| 0.0% | 5.0% | Initiate | 0 (0%) | 155,707 (3%) | | 832,684 (17%) | 1,669,220 (33%) | 2,499,448 (50%) | 671,995 (13%) | 87,629 (2%) | 4,241,728 (85%) |
| | | Not initiate | 0 (0%) | 231,159 (5%) | | 2,497,575 (50%) | 1,667,618 (33%) | 833,455 (17%) | 888,086 (18%) | 67,555 (1%) | 4,043,007 (81%) |
| 2.5% | 2.5% | Initiate | 437,406 (9%) | 77,732 (2%) | | 833,579 (17%) | 1,668,628 (33%) | 2,499,270 (50%) | 672,163 (13%) | 87,607 (2%) | 4,241,707 (85%) |
| | | Not initiate | 185,860 (4%) | 117,857 (2%) | | 2,496,680 (50%) | 1,668,210 (33%) | 833,633 (17%) | 888,050 (18%) | 67,433 (1%) | 4,043,040 (81%) |
| 5.0% | 0.0% | Initiate | 787,781 (16%) | 0 (0%) | | 830,772 (17%) | 1,668,317 (33%) | 2,499,639 (50%) | 672,553 (13%) | 87,581 (2%) | 4,238,594 (85%) |
| | | Not initiate | 344,591 (7%) | 0 (0%) | | 2,499,487 (50%) | 1,668,521 (33%) | 833,264 (17%) | 887,123 (18%) | 67,164 (1%) | 4,046,985 (81%) |

**Initiation does not affect the risk of abortion or preeclampsia.**

| | | | | | | | | | | | |
|---|---|---|---|---|---|---|---|---|---|---|---|
| 0.0% | 5.0% | Initiate | 0 (0%) | 288,440 (6%) | 1 | 833,846 (17%) | 1,666,311 (33%) | 2,500,877 (50%) | 1,136,208 (23%) | 78,421 (2%) | 3,786,405 (76%) |
| | | Not initiate | 0 (0%) | 230,079 (5%) | 0 | 2,499,144 (50%) | 1,666,083 (33%) | 833,739 (17%) | 888,982 (18%) | 67,949 (1%) | 4,042,035 (81%) |
| 2.5% | 2.5% | Initiate | 393,597 (8%) | 146,457 (3%) | 1 | 832,531 (17%) | 1,665,783 (33%) | 2,500,978 (50%) | 1,136,774 (23%) | 78,129 (2%) | 3,784,389 (76%) |
| | | Not initiate | 186,215 (4%) | 117,538 (2%) | 0 | 2,500,459 (50%) | 1,666,611 (33%) | 833,638 (17%) | 889,004 (18%) | 67,602 (1%) | 4,044,102 (81%) |
| 5.0% | 0.0% | Initiate | 711,240 (14%) | 0 (0%) | 1 | 832,811 (17%) | 1,665,041 (33%) | 2,500,013 (50%) | 1,136,167 (23%) | 78,097 (2%) | 3,783,601 (76%) |
| | | Not initiate | 345,016 (7%) | 0 (0%) | 0 | 2,500,179 (50%) | 1,667,353 (33%) | 834,603 (17%) | 890,226 (18%) | 67,771 (1%) | 4,044,138 (81%) |

**Initiation increases the risk of abortion and does not affect the risk of preeclampsia.**

| | | | | | | | | | | |
|---|---|---|---|---|---|---|---|---|---|---|
| 0.0% | 5.0% | Initiate | 0 (0%) | 582,391 (12%) | 834,008 (17%) | 1,666,954 (33%) | 2,499,411 (50%) | 2,084,468 (42%) | 61,846 (1%) | 2,854,059 (57%) |
| | | Not initiate | 0 (0%) | 230,422 (5%) | 2,499,619 (50%) | 1,666,685 (33%) | 833,323 (17%) | 889,519 (18%) | 67,580 (1%) | 4,042,528 (81%) |
| 2.5% | 2.5% | Initiate | 340,955 (7%) | 300,202 (6%) | 834,174 (17%) | 1,667,025 (33%) | 2,499,190 (50%) | 2,085,160 (42%) | 61,859 (1%) | 2,853,370 (57%) |
| | | Not initiate | 186,248 (4%) | 118,138 (2%) | 2,499,453 (50%) | 1,666,614 (33%) | 833,544 (17%) | 889,747 (18%) | 67,350 (1%) | 4,042,514 (81%) |
| 5.0% | 0.0% | Initiate | 615,364 (12%) | 0 (0%) | 834,002 (17%) | 1,665,188 (33%) | 2,500,448 (50%) | 2,084,425 (42%) | 62,192 (1%) | 2,853,021 (57%) |
| | | Not initiate | 344,553 (7%) | 0 (0%) | 2,499,625 (50%) | 1,668,451 (33%) | 832,286 (17%) | 889,344 (18%) | 67,564 (1%) | 4,043,454 (81%) |

[a] Defined as fetal death prior to 20 weeks of gestation from LMP (18 from conception).
[b] Defined as fetal death at or after 20 weeks of gestation from LMP (18 from conception).

**Table S9.** Descriptive statistics regarding the number of pregnancies and prenatal preeclampsia outcomes included in each analytic sample where ~20% of pregnancies were missing outcomes by treatment.

| Goal Percentage Missing Due to… | | | Target Population | | Analytic Sample | | | | | |
|---|---|---|---|---|---|---|---|---|---|---|
| | | | | | Observed Deliveries | | Observed Outcomes | | Observed Pregnancies | |
| Severity or Rurality | Miscarriage | Treatment | Pregnancies | Preeclampsia | Pregnancies | Preeclampsia | Pregnancies | Preeclampsia | Pregnancies | Preeclampsia |
| *Initiation decreases the risk of miscarriage and preeclampsia.* | | | | | | | | | | |
| 0.0% | 20.0% | Initiate | 5,000,779 | 1,626,321 | 4,328,048 | 1,626,321 | 4,347,837 | 1,626,321 | 5,000,779 | 1,626,321 |
| | | Not initiate | 4,999,221 | 1,760,727 | 4,112,000 | 1,760,727 | 4,133,031 | 1,760,727 | 4,999,221 | 1,760,727 |
| 10.0% | 10.0% | Initiate | 5,000,809 | 1,625,447 | 3,057,861 | 1,019,944 | 3,382,415 | 1,019,944 | 5,000,809 | 1,625,447 |
| | | Not initiate | 4,999,191 | 1,760,483 | 3,531,920 | 1,464,064 | 3,962,134 | 1,464,064 | 4,999,191 | 1,760,483 |
| 20.0% | 0.0% | Initiate | 4,997,555 | 1,624,608 | 2,382,840 | 719,242 | 2,933,751 | 719,242 | 4,997,555 | 1,624,608 |
| | | Not initiate | 5,002,445 | 1,761,592 | 3,145,819 | 1,274,840 | 3,971,567 | 1,274,840 | 5,002,445 | 1,761,592 |
| *Initiation does not affect the risk of miscarriage and decreases the risk of preeclampsia.* | | | | | | | | | | |
| 0.0% | 20.0% | Initiate | 4,997,204 | 1,459,920 | 3,860,925 | 1,459,920 | 3,888,993 | 1,459,920 | 4,997,204 | 1,459,920 |
| | | Not initiate | 5,002,796 | 1,761,806 | 4,113,673 | 1,761,806 | 4,135,040 | 1,761,806 | 5,002,796 | 1,761,806 |
| 10.0% | 10.0% | Initiate | 4,999,263 | 1,459,594 | 2,732,335 | 919,764 | 3,265,957 | 919,764 | 4,999,263 | 1,459,594 |
| | | Not initiate | 5,000,737 | 1,761,711 | 3,534,057 | 1,464,637 | 3,962,955 | 1,464,637 | 5,000,737 | 1,761,711 |
| 20.0% | 0.0% | Initiate | 5,000,339 | 1,459,901 | 2,129,354 | 649,152 | 3,114,170 | 649,152 | 5,000,339 | 1,459,901 |
| | | Not initiate | 4,999,661 | 1,761,445 | 3,145,981 | 1,275,161 | 3,971,067 | 1,275,161 | 4,999,661 | 1,761,445 |
| *Initiation increases the risk of miscarriage and decreases the risk of preeclampsia.* | | | | | | | | | | |
| 0.0% | 20.0% | Initiate | 5,000,353 | 1,210,493 | 2,919,099 | 1,210,493 | 2,961,457 | 1,210,493 | 5,000,353 | 1,210,493 |
| | | Not initiate | 4,999,647 | 1,759,873 | 4,109,846 | 1,759,873 | 4,131,190 | 1,759,873 | 4,999,647 | 1,759,873 |
| 10.0% | 10.0% | Initiate | 5,001,451 | 1,212,062 | 1,964,615 | 744,743 | 2,905,992 | 744,743 | 5,001,451 | 1,212,062 |
| | | Not initiate | 4,998,549 | 1,758,730 | 3,532,992 | 1,462,629 | 3,962,921 | 1,462,629 | 4,998,549 | 1,758,730 |
| 20.0% | 0.0% | Initiate | 5,000,608 | 1,211,427 | 1,467,873 | 511,671 | 3,373,389 | 511,671 | 5,000,608 | 1,211,427 |
| | | Not initiate | 4,999,392 | 1,759,151 | 3,144,756 | 1,274,211 | 3,971,533 | 1,274,211 | 4,999,392 | 1,759,151 |

**Initiation decreases the risk of abortion and does not affect the risk of preeclampsia.**

| | | | | | | | | | | | |
|---|---|---|---|---|---|---|---|---|---|---|---|
| 0.0% | 20.0% | | Initiate | 5,002,131 | 2,217,135 | 4,330,383 | 2,217,135 | 4,349,808 | 2,217,135 | 5,002,131 | 2,217,135 |
| | | | Not initiate | 4,997,869 | 1,760,053 | 4,109,530 | 1,760,053 | 4,130,687 | 1,760,053 | 4,997,869 | 1,760,053 |
| 10.0% | 10.0% | | Initiate | 4,998,871 | 2,215,956 | 3,073,925 | 1,512,160 | 3,399,345 | 1,512,160 | 4,998,871 | 2,215,956 |
| | | | Not initiate | 5,001,129 | 1,761,095 | 3,533,463 | 1,465,021 | 3,962,802 | 1,465,021 | 5,001,129 | 1,761,095 |
| 20.0% | 0.0% | | Initiate | 5,002,776 | 2,217,528 | 2,406,228 | 1,145,032 | 2,957,270 | 1,145,032 | 5,002,776 | 2,217,528 |
| | | | Not initiate | 4,997,224 | 1,759,968 | 3,142,015 | 1,274,057 | 3,968,022 | 1,274,057 | 4,997,224 | 1,759,968 |

**Initiation does not affect the risk of abortion or preeclampsia.**

| | | | | | | | | | | | |
|---|---|---|---|---|---|---|---|---|---|---|---|
| 0.0% | 20.0% | | Initiate | 5,000,810 | 1,982,196 | 3,864,132 | 1,982,196 | 3,892,544 | 1,982,196 | 5,000,810 | 1,982,196 |
| | | | Not initiate | 4,999,190 | 1,759,174 | 4,109,145 | 1,759,174 | 4,130,175 | 1,759,174 | 4,999,190 | 1,759,174 |
| 10.0% | 10.0% | | Initiate | 5,002,326 | 1,983,319 | 2,745,635 | 1,355,288 | 3,280,099 | 1,355,288 | 5,002,326 | 1,983,319 |
| | | | Not initiate | 4,997,674 | 1,758,948 | 3,530,703 | 1,462,890 | 3,960,683 | 1,462,890 | 4,997,674 | 1,758,948 |
| 20.0% | 0.0% | | Initiate | 5,000,594 | 1,982,744 | 2,146,179 | 1,024,019 | 3,133,220 | 1,024,019 | 5,000,594 | 1,982,744 |
| | | | Not initiate | 4,999,406 | 1,759,089 | 3,144,121 | 1,273,871 | 3,971,229 | 1,273,871 | 4,999,406 | 1,759,089 |

**Initiation increases the risk of abortion and does not affect the risk of preeclampsia.**

| | | | | | | | | | | | |
|---|---|---|---|---|---|---|---|---|---|---|---|
| 0.0% | 20.0% | | Initiate | 4,997,123 | 1,558,351 | 2,913,685 | 1,558,351 | 2,956,341 | 1,558,351 | 4,997,123 | 1,558,351 |
| | | | Not initiate | 5,002,877 | 1,759,883 | 4,113,265 | 1,759,883 | 4,134,295 | 1,759,883 | 5,002,877 | 1,759,883 |
| 10.0% | 10.0% | | Initiate | 4,998,686 | 1,559,647 | 1,974,777 | 1,023,485 | 2,913,383 | 1,023,485 | 4,998,686 | 1,559,647 |
| | | | Not initiate | 5,001,314 | 1,760,373 | 3,532,743 | 1,463,872 | 3,962,744 | 1,463,872 | 5,001,314 | 1,760,373 |
| 20.0% | 0.0% | | Initiate | 4,999,076 | 1,560,040 | 1,482,661 | 746,935 | 3,389,010 | 746,935 | 4,999,076 | 1,560,040 |
| | | | Not initiate | 5,000,924 | 1,758,929 | 3,143,356 | 1,273,913 | 3,970,854 | 1,273,913 | 5,000,924 | 1,758,929 |

**Table S10.** Descriptive statistics regarding the number of pregnancies and prenatal preeclampsia outcomes included in each analytic sample where ~5% of pregnancies were missing outcomes by treatment.

| Goal Percentage Missing Due to… | | | Study Sample | | Analytic Sample | | | | | |
|---|---|---|---|---|---|---|---|---|---|---|
| | | | | | Observed Deliveries | | Observed Outcomes | | Observed Pregnancies | |
| Severity or Rurality | Miscarriage | Treatment | Pregnancies | Preeclampsia | Pregnancies | Preeclampsia | Pregnancies | Preeclampsia | Pregnancies | Preeclampsia |
| *Initiation decreases the risk of miscarriage and preeclampsia.* | | | | | | | | | | |
| 0.0% | 5.0% | Initiate | 5,000,373 | 1,558,664 | 2,915,905 | 1,558,664 | 4,417,982 | 1,558,664 | 5,000,373 | 1,558,664 |
| | | Not initiate | 4,999,627 | 1,759,228 | 4,110,108 | 1,759,228 | 4,769,205 | 1,759,228 | 4,999,627 | 1,759,228 |
| 2.5% | 2.5% | Initiate | 5,000,389 | 1,559,298 | 2,598,947 | 1,379,071 | 4,359,232 | 1,379,071 | 5,000,389 | 1,559,298 |
| | | Not initiate | 4,999,611 | 1,759,062 | 3,932,302 | 1,666,212 | 4,695,225 | 1,666,212 | 4,999,611 | 1,759,062 |
| 5.0% | 0.0% | Initiate | 4,999,638 | 1,559,062 | 2,348,312 | 1,236,320 | 4,384,274 | 1,236,320 | 4,999,638 | 1,559,062 |
| | | Not initiate | 5,000,362 | 1,760,007 | 3,783,381 | 1,589,744 | 4,655,809 | 1,589,744 | 5,000,362 | 1,760,007 |
| *Initiation does not affect the risk of miscarriage and decreases the risk of preeclampsia.* | | | | | | | | | | |
| 0.0% | 5.0% | Initiate | 4,998,846 | 1,458,858 | 3,862,447 | 1,458,858 | 4,709,913 | 1,458,858 | 4,998,846 | 1,458,858 |
| | | Not initiate | 5,001,154 | 1,761,505 | 4,113,047 | 1,761,505 | 4,770,640 | 1,761,505 | 5,001,154 | 1,761,505 |
| 2.5% | 2.5% | Initiate | 5,001,503 | 1,460,622 | 3,486,755 | 1,277,400 | 4,454,927 | 1,277,400 | 5,001,503 | 1,460,622 |
| | | Not initiate | 4,998,497 | 1,761,128 | 3,934,049 | 1,668,475 | 4,694,928 | 1,668,475 | 4,998,497 | 1,761,128 |
| 5.0% | 0.0% | Initiate | 4,998,053 | 1,459,356 | 3,183,369 | 1,132,171 | 4,276,872 | 1,132,171 | 4,998,053 | 1,459,356 |
| | | Not initiate | 5,001,947 | 1,761,958 | 3,784,640 | 1,591,533 | 4,656,702 | 1,591,533 | 5,001,947 | 1,761,958 |
| *Initiation increases the risk of miscarriage and decreases the risk of preeclampsia.* | | | | | | | | | | |
| 0.0% | 5.0% | Initiate | 4,999,781 | 1,211,595 | 2,919,337 | 1,211,595 | 4,419,968 | 1,211,595 | 4,999,781 | 1,211,595 |
| | | Not initiate | 5,000,219 | 1,759,505 | 4,111,549 | 1,759,505 | 4,769,451 | 1,759,505 | 5,000,219 | 1,759,505 |
| 2.5% | 2.5% | Initiate | 4,999,251 | 1,211,119 | 2,597,410 | 1,052,315 | 4,354,627 | 1,052,315 | 4,999,251 | 1,211,119 |
| | | Not initiate | 5,000,749 | 1,760,652 | 3,935,501 | 1,667,810 | 4,695,925 | 1,667,810 | 5,000,749 | 1,760,652 |
| 5.0% | 0.0% | Initiate | 5,002,459 | 1,212,360 | 2,345,887 | 929,128 | 4,378,385 | 929,128 | 5,002,459 | 1,212,360 |
| | | Not initiate | 4,997,541 | 1,758,074 | 3,780,908 | 1,588,610 | 4,653,655 | 1,588,610 | 4,997,541 | 1,758,074 |
| *Initiation decreases the risk of abortion and does not affect the risk of preeclampsia.* | | | | | | | | | | |

| | | | | | | | | | | | |
|---|---|---|---|---|---|---|---|---|---|---|---|
| 0.0% | 5.0% | Initiate | 5,001,352 | 2,217,834 | 4,329,357 | 2,217,834 | 4,845,645 | 2,217,834 | 5,001,352 | 2,217,834 |
| | | Not initiate | 4,998,648 | 1,760,807 | 4,110,562 | 1,760,807 | 4,767,489 | 1,760,807 | 4,998,648 | 1,760,807 |
| 2.5% | 2.5% | Initiate | 5,001,477 | 2,217,874 | 3,909,814 | 1,981,682 | 4,486,339 | 1,981,682 | 5,001,477 | 2,217,874 |
| | | Not initiate | 4,998,523 | 1,760,967 | 3,933,362 | 1,668,200 | 4,694,806 | 1,668,200 | 4,998,523 | 1,760,967 |
| 5.0% | 0.0% | Initiate | 4,998,728 | 2,215,729 | 3,573,027 | 1,791,921 | 4,210,947 | 1,791,921 | 4,998,728 | 2,215,729 |
| | | Not initiate | 5,001,272 | 1,762,644 | 3,786,657 | 1,592,615 | 4,656,681 | 1,592,615 | 5,001,272 | 1,762,644 |

**Initiation does not affect the risk of abortion or preeclampsia.**

| | | | | | | | | | | | |
|---|---|---|---|---|---|---|---|---|---|---|---|
| 0.0% | 5.0% | Initiate | 5,001,034 | 1,983,310 | 3,864,826 | 1,983,310 | 4,712,594 | 1,983,310 | 5,001,034 | 1,983,310 |
| | | Not initiate | 4,998,966 | 1,760,105 | 4,109,984 | 1,760,105 | 4,768,887 | 1,760,105 | 4,998,966 | 1,760,105 |
| 2.5% | 2.5% | Initiate | 4,999,292 | 1,981,973 | 3,490,526 | 1,772,618 | 4,459,238 | 1,772,618 | 4,999,292 | 1,981,973 |
| | | Not initiate | 5,000,708 | 1,761,117 | 3,934,342 | 1,668,059 | 4,696,955 | 1,668,059 | 5,000,708 | 1,761,117 |
| 5.0% | 0.0% | Initiate | 4,997,865 | 1,981,827 | 3,192,409 | 1,604,906 | 4,286,625 | 1,604,906 | 4,997,865 | 1,981,827 |
| | | Not initiate | 5,002,135 | 1,760,621 | 3,784,122 | 1,590,950 | 4,657,119 | 1,590,950 | 5,002,135 | 1,760,621 |

**Initiation increases the risk of abortion and does not affect the risk of preeclampsia.**

| | | | | | | | | | | | |
|---|---|---|---|---|---|---|---|---|---|---|---|
| 0.0% | 5.0% | Initiate | 2,497,775 | 779,021 | 1,456,940 | 779,021 | 2,207,337 | 779,021 | 2,497,775 | 779,021 |
| | | Not initiate | 2,502,225 | 880,572 | 2,057,827 | 880,572 | 2,387,422 | 880,572 | 2,502,225 | 880,572 |
| 2.5% | 2.5% | Initiate | 2,499,859 | 780,547 | 1,300,572 | 690,621 | 2,179,949 | 690,621 | 2,499,859 | 780,547 |
| | | Not initiate | 2,500,141 | 880,001 | 1,967,156 | 833,593 | 2,347,924 | 833,593 | 2,500,141 | 880,001 |
| 5.0% | 0.0% | Initiate | 2,501,130 | 781,649 | 1,175,850 | 619,724 | 2,193,340 | 619,724 | 2,501,130 | 781,649 |
| | | Not initiate | 2,498,870 | 878,581 | 1,890,915 | 794,058 | 2,326,474 | 794,058 | 2,498,870 | 878,581 |